# Surface-guided computing to analyze subcellular morphology and membrane-associated signals in 3D

**Authors**
Felix Y. Zhou[1,2], Andrew Weems[1,2], Gabriel M. Gihana[1,2], Bingying Chen[1,2], Bo-Jui Chang[1,2], Meghan Driscoll[1,2,3] and Gaudenz Danuser[1,2]

**Affiliation**
[1]Lyda Hill Department of Bioinformatics, University of Texas Southwestern Medical Center, Dallas, TX, USA.
[2]Cecil H. & Ida Green Center for System Biology, University of Texas Southwestern Medical Center, Dallas, TX, USA
[3]Current address: Department of Pharmacology, University of Minnesota, Minneapolis, MN, USA

**Correspondence**
Correspondence to Felix Zhou or Gaudenz Danuser: {felix.zhou, gaudenz.danuser}@utsouthwestern.edu

**Abstract**
Signal transduction and cell function are governed by the spatiotemporal organization of membrane-associated molecules. Despite significant advances in visualizing molecular distributions by 3D light microscopy, cell biologists still have limited quantitative understanding of the processes implicated in the regulation of molecular signals at the whole cell scale. In particular, complex and transient cell surface morphologies challenge the complete sampling of cell geometry, membrane-associated molecular concentration and activity and the computing of meaningful parameters such as the cofluctuation between morphology and signals. Here, we introduce u-Unwrap3D, a framework to remap arbitrarily complex 3D cell surfaces and membrane-associated signals into equivalent lower dimensional representations. The mappings are bidirectional, allowing the application of image processing operations in the data representation best suited for the task and to subsequently present the results in any of the other representations, including the original 3D cell surface. Leveraging this surface-guided computing paradigm, we track segmented surface motifs in 2D to quantify the recruitment of Septin polymers by blebbing events; we quantify actin enrichment in peripheral ruffles; and we measure the speed of ruffle movement along topographically complex cell surfaces. Thus, u-Unwrap3D provides access to spatiotemporal analyses of cell biological parameters on unconstrained 3D surface geometries and signals.

**Main**

Advances in 3D high-resolution live-cell microscopy and biosensor design enable integrative studies of the dynamic interplay and causal relations between cell morphology and signal transduction *in vitro* and *in vivo*. By reshaping the plasma membrane into diverse morphologies, cells sense, respond to and remodel their local environment[1-6]. Many cell types adopt shapes that are tailored to their characteristic function[7-11]. Cell morphology has thus long been recognised as a proxy of cell state and as a marker of differentiation[9,11,12]. Mechanically, the plasma membrane integrates internal and external forces, which affects cell fate through mechanotransductive proteins and changes in cytoskeleton and nuclear morphology[13-15]. Structurally, the plasma membrane serves as a platform for catalysing chemical reactions[16-20] and as a spatiotemporal organiser of signalling activity through the creation of binding sites, local confinements and molecular concentration in scaffolds, diffusion traps, and by phase separation[16,21-23]. These reactions occur locally at the nanometer or micron length scale or in global bursts that span the entire cell[18,24]. Understanding the salient biophysical processes that govern the formation and persistence of these subcellular signalling domains and how these domains regulate biochemical signal transduction remains enigmatic. Systematic identification of this intricate regulatory interplay between cell shape and molecular signalling necessitates consistent temporal tracking of the local 3D cell geometry and conjoint sampling of the corresponding membrane-associated molecular concentration and activity.

Cell surfaces are extracted from binary segmented image volumes and stored as a mesh, a data structure described by a list of the Cartesian 3D vertex coordinates on the surface and a second list specifying how the individual vertices are connected into triangles or faces. Tracking the correspondence between two 3D surface meshes is an active area of research in computer graphics[25-28] and none of the methods have been adopted to cell imaging. A particular technical challenge that arises when adapting techniques from computer graphics with applications to cell biology is the non-convexity, irregularity and high curvature of surface protrusions on most cell shapes. Very few methods have been proposed to accurately follow such geometries over time and have largely been demonstrated on well-defined shapes such as human pose[29] or hands[30,31]. Generally, these methods track by matching meshes from consecutive timepoints. To match meshes, methods attempt to assign a unique signature per vertex or face to establish a matching between vertices and faces by minimizing a loss metric[32,33]. However, this approach is inherently sensitive to mesh quality, uniqueness of the signature, optimizer convergence and is difficult to generalize when tracking surfaces over many timepoints. Crucially, meshes segmented from two different timepoints have different numbers of vertices and faces and the lack of the exact same surface features poses ambiguity in matching. Alternatively, individual 3D image volumes may be first registered spatiotemporally before mesh extraction, as is done in neuroscience[34-37]. This approach is robust to mesh quality and sampling errors, however deformations must be small between timepoints. For example, long or thin cell surface structures such as lamellipodia and filopodia suffer voxel undersampling, limiting the registration to cell surfaces with largely globular features such as blebs[38]. This problem could be remedied by mapping the 3D surface to the unit sphere[39-41]. Indeed, in macroscopic imaging application, this procedure has enabled registration of complex geometrical features such as brain folds directly on the 3D sphere or in derivative 2D unwrapped images[27,42-44]. Unfortunately, this strategy requires closed surface topologies with no holes (genus-0 surfaces), which is not generally guaranteed in live-cell microscopy. Alternatively, one can selectively segment surface motifs and track these in 3D whilst mapping surface-proximal molecular signal intensity[38,45-47]. Like the 3D surface tracking, this approach is also susceptible to the variable quality of the segmented motifs used for matching and non-convexities of the surface.

Here, we develop a general and comprehensive software solution, u-Unwrap3D, for surface-guided computing. u-Unwrap3D remaps arbitrarily complex 3D subcellular morphology and membrane associated signals to equivalent lower dimensional representations that allow for optimized computation of surface features and spatiotemporal tracking and sampling of cell geometry and associated molecular entities. We demonstrate the power of this approach in applications to i) the unsupervised segmentation of diverse surface motifs; ii) the quantification of septin polymer recruitment to dynamic cell surface blebs and iii) the measurement of travel speed of actin-enriched surface ruffles.

**Results**

**u-Unwrap3D for surface-guided computing**

Given an input Cartesian 3D surface $S(x,y,z)$ as a 3D mesh, u-Unwrap3D computes a series of equivalent surface representations (Fig. 1a, Suppl. Video 1). The input surface $S(x,y,z)$ is smoothened to find a genus-0 reference surface $S_{\text{ref}}(x,y,z)$ and then mapped to the sphere, $S^2(x,y,z)$, unwrapped into a 2D $S_{\text{ref}}(u,v)$ using UV-unwrapping and the $(u,v)$ parameterized $S_{\text{ref}}(x,y,z)$ propagated along the steepest gradient of its signed distance transform to construct a topographic representation of the input surface, $S(d,u,v)$. The mappings between representations are bijective and constructed to minimize the associated conformal (preservation of aspect ratio) and equiareal (preservation of surface area fraction) errors[48] (Extended Fig. 1a,b). We denote $S(\cdot)$ and $V(\cdot)$ as surfaces and volumes, respectively, relative to a coordinate system indicated in brackets. The variables $F_i(S(\cdot))$ and $I_i(V(\cdot))$ denote surface- and volume- associated signals of interest. These signals may describe geometrical quantities, (like mean curvature, *H*), integer labels (like segmented surface protrusions) or molecular activities (like molecular concentrations or activities). Thus, u-Unwrap3D provides a framework to map these variables between different surface representations, each of which is suited for different computational tasks.

**Step 1** iteratively smoothens out salient surface features on $S(x,y,z)$ using conformalized mean curvature flow (cMCF)[49] to determine a genus-0 reference surface $S_{\text{ref}}(x,y,z)$ without holes or 'handles'. The term 'handle' refers to the holes in a loop mesh structure such as the handle of a teacup that unlike 'holes' does not involve missing/incomplete surface patches in a mesh. cMCF iteratively displaces vertices with a speed proportional to the mean curvature at each vertex (Extended Fig. 1c). The input surface is thereby preferentially deformed into the largest inscribable sphere. In the absence of *a priori* markers for the reference shape such as cell cortex markers, the rate of decrease in mean absolute Gaussian curvature $K$ is monitored to determine a stopping iteration (Methods). The Gaussian curvature $K$ is a shape-invariant measure of local curvature. Accordingly, the same shape (e.g. a sphere) has identical $K$ value irrespective of size[50]. $K$ is thus well-suited as a criterion to terminate the cMCF iterations. cMCF is agnostic to minor mesh imperfections such as small holes and handles but does not change the genus. Any holes or handles in the input surface are still present. However, because of the smoothing, these holes are more regular and smaller. Still in Step 1, we compute a genus-0 mesh of the reference surface $S_{\text{ref}}(x,y,z)$ through filling all holes in the volume enclosed by the intermediary cMCF-processed surface and remeshing of the resulting body (Methods). The remeshing changes the vertex position and face topology. To restore bijectivity between the input surface $S(x,y,z)$ and $S_{\text{ref}}(x,y,z)$ we match the mesh $S_{\text{ref}}(x,y,z)$ with the mesh of the intermediary cMCF-processed surface that is bijective to $S(x,y,z)$, (Methods). Any associated measurements $F_i(S(x,y,z))$ are mapped to $F_i(S_{\text{ref}}(x,y,z))$ by interpolation. In **Step 2**, the genus-0 reference surface, $(S_{\text{ref}}(x,y,z)$ is quasi-conformally mapped to the unit sphere without folds[39,51] (Extended Fig. 1d). This spherical parametrization is denoted $S_Q^2(x,y,z)$. Per the uniformization theorem, such a mapping always exists for a genus-0 surface[52-55]. The quasi-conformal spherical parameterization $S_Q^2(x,y,z)$ severely shrinks surface extremities deviating from the sphere[27], even for roughly globular shapes (Extended Fig. 1d). Consequently, surface features with high curvature are undersampled and disproportionately represented relative to their original Cartesian 3D surface area. This can detrimentally affect downstream analyses such as segmentation and tracking[27]. To mitigate this problem, we iteratively diffuse in **Step 3** the area distortion factor per face by advecting vertex positions on the sphere[27] at the expense of increased conformal error (Extended Fig. 1e). In **Step 4,** this quasi-equiareal sphere $S_\Omega^2(x,y,z)$ is bijectively unwrapped to the 2D plane, $S(u,v)$, using equirectangular projection, in short UV-mapping, with $(u,v)$ denoting the spherical polar and azimuthal angles, respectively. UV-mapping introduces the strongest distortions to signals at the north and south poles of the sphere. To visualize features of interest with minimal distortion, u-Unwrap3D optionally infers a rotation matrix based on a weighted principal component analysis of surface variables, such as the local curvature (Extended Fig. 1f,g, Methods). If the input Cartesian 3D surface mesh $S(x,y,z)$ is genus-0, the generation of a reference surface, $S_{\text{ref}}(x,y,z)$ may be skipped and 2D equiareal surface unwrapping realised directly (Fig. 1b, Suppl. Video 2). We note that an input genus-X Cartesian 3D surface $S(x,y,z)$ is also directly unwrapped into 2D through steps 1-4 of u-Unwrap3D, but not in an equiareal manner (Extended Fig. 1h). In **Step 5**, the first part remaps the Cartesian 3D volume $V(x,y,z)$ and associated signals $I_i(V(x,y,z))$ into a topographic volume $V(d,u,v)$ coordinate system that is normal to the reference surface, $S_{\text{ref}}(x,y,z)$. The second part establishes a bijective mapping of $(d,u,v)$ to $(x,y,z)$ coordinates, $S_{\text{ref}}(u,v)$, i.e. the $(u,v)$ parameterized reference surface of $S_{\text{ref}}(x,y,z)$, is propagated in Cartesian 3D space in the surface normal direction at equidistant steps of $\alpha$ voxels along the steepest gradient of the signed distance function, $\nabla\Phi(x,y,z)$, for a total of $D$ steps. Interpolation of the respective Cartesian volumetric signal intensities, $I_i(V(x,y,z))$ at the $(x,y,z)$ coordinates indexed by $(d,u,v)$ generates the topographic 3D volume equivalents, $I_i(V(d,u,v))$. Finally, in **Step 6** the topographic 3D surface representation, $S(d,u,v)$ of the input surface $S(x,y,z)$ is obtained by surface meshing the topographic binary volume segmentation.

In summary, u-Unwrap3D generates bijective mappings of a given genus-X surface between 5 equivalent surface representations; Cartesian 3D, $S(x,y,z)$, genus-0 reference 3D, $S_{\text{ref}}(x,y,z)$, the unit 3D sphere, $S^2(x,y,z)$, topographic 3D, $S(d,u,v)$, and the 2D plane image, $S(u,v)$ (Fig. 2a), while simultaneously transforming Cartesian 3D to topographic 3D volumes (Fig. 2b). This was made possible by two crucial choices; the use of cMCF and voxelization to construct a genus-0 reference surface, $S_{\text{ref}}(x,y,z)$ to realise spherical parameterization (Step 1) and the implementation of an efficient numerical scheme to relax area distortion on the 3D sphere (Step 3). The former allows us to construct $S_{\text{ref}}(x,y,z)$ as a proxy of the genus-X $S(x,y,z)$ surface mesh and to unwrap this 3D surface into one 2D $(u,v)$ image, instead of requiring multiple 2D $(u,v)$ images, which simplifies downstream analysis[56-59]. The latter ensures that the unwrapped 2D $(u,v)$ image captures the salient surface features of $S_{\text{ref}}(x,y,z)$, and by extension the genus-X $S(x,y,z)$ surface. Importantly, the bijectivity of the mappings guarantees that for any point on any of the surface or volume representations matching points exist on any of the other surfaces or volumes. Moreover, the bijectivity guarantees preservation of the point topology, i.e. a series of points ordered in clockwise fashion on one surface representation maps to a series of points ordered in the same way on any of the other surface representations and preserves the local neighbourhood relationships. As a result, we can apply mathematical operations defined in any one of the representations and map the results to any other. u-Unwrap3D thus supports the optimal spatiotemporal analysis of unconstrained surface geometries and associated signals.

**Validation of u-Unwrap3D on diverse surface motifs**
We validated the generality and performance of u-Unwrap3D by application to 66 single cell images acquired by high-resolution light sheet imaging[45,60]. The dataset span morphologically diverse cells with blebs, lamellipodia and filopodia. The cell surfaces were meshed with marching cubes and segmented within the u-Shape3D software[45]. Small errors in the initial segmentation and meshing process cause high-order genus surfaces with topological holes and handles, which cannot be unwrapped directly (Extended Fig. 2a). Holes can also generate non-watertight surface meshes – surfaces that are not closed and have no clearly defined inside volume[48,61] possessing potentially complex internal volumetric structures that violate the assumptions of standard 3D mesh processing algorithms.

We first tested the number of input cell surfaces for which u-Unwrap3D could successfully run all steps 1-6 and compute all 5 of the representations as a measure of generality and robustness. Notably only 6/66 (11%) input cell surfaces were genus-0 and only 36/66 (55%) were watertight (Extended Fig 2b). In 63/66 cases, (>95%) we successfully ran all steps and obtained all representations (Extended Fig 2b). The three failures occurred in scenarios, in which the holes and handles remaining after the application of cMCF were still too large for the volume dilation to generate a genus-0 reference surface after remeshing (Fig. 1b, Step 1) (Extended Fig. 2c). In all successful cases, cMCF and binary voxelization under volume dilation generated genus-0 reference surfaces within a median of 10 iterations (Extended Fig 2c, c.f. lamellipodia). Fig. 2c shows extracted representations for challenging examples with blebs, lamellipodia and filopodia (Suppl. Video 3-5).

We next tested the robustness and performance of the $S_{\text{ref}}(x,y,z)$ spherical parameterizations, (Fig. 1b, Steps 2-3). Extended Fig. 3a confirms that the quasi-conformal spherical parameterization (Step 2) minimizes the conformal error to the ideal value of 1, with the largest error in cells with filopodia ($1.016\pm0.013$). We also verified the need to relax local area distortion. Whilst quasi-conformal spherical parameterization $S_\Omega^2(x,y,z)$ is equiareal for blebs ($0.978\pm0.037$), the median area distortion showed that the surface fraction of lamellipodia was down to 0.432 and in filopodia to just 0.140 with respect to their original area fraction on the reference 3D, $S_{\text{ref}}(x,y,z)$ surface, let alone $S(x,y,z)$. Our scheme for area distortion relaxation (Methods) produces a quasi-equiareal spherical parameterization (Step 3) in blebs (0.985±0.012), and successfully achieves the ideal value of 1 in lamellipodia (1.000±0.000) and filopodia(1.000±0.001), within a maximum median of 23 iterations for lamellipodia (Extended Fig.3b, Table i). We further tested the ability of our relaxation scheme to balance the trade-off between the two extremes of conformal to equiareal spherical parameterizations using different stopping criteria (Extended Fig.3b, Table ii-iv). The initial parameterization without any area-distortion relaxation (iteration 0) is by design conformal but also found to satisfy the most isometric parameterization (MIP)[62]. Running for $t_\Omega <$ a maximum of 50 iterations yields an equiareal parameterization for all motifs. At $t \approx \frac{1}{2} t_\Omega$ iterations the relaxed mesh jointly minimizes the summation ($Q +$ $\ln \lambda$, Methods) of conformal ($Q$) and area distortion ($\lambda$) errors. At $t \lesssim t_\Omega$ iterations the relaxed mesh is the area-preserving MIP[63]. As expected, this latter parameterization does not fully minimize area distortion (blebs (0.997±0.011), lamellipodia (0.979±0.005) and filopodia (0.980±0.028)) but exhibits slightly lower conformal errors and consequently higher quality faces than a pure equiareal mapping.

Lastly, we tested how accurately $S_{\text{topo}}(x,y,z)$, which defines the topographic 3D mesh, $S(d,u,v)$ (step 5) mapped back into Cartesian coordinates reconstructs the input surface, $S(x,y,z)$. For all cells, $S(d,u,v)$ was computed with a $(u,v)$ image grid size of 1024x512 pixels. The aspect ratio, $2N$ x $N$ ($N = 512$) was chosen to preserve the ratio between the equatorial circumference and the length of the arc between north and south poles of a sphere. Compared to the input surface $S(x,y,z)$, $S(d,u,v)$ is lower genus and provides higher face quality (Extended Fig. 3c). We assessed the discrepancy between $S_{\text{topo}}(x,y,z)$ and $S(x,y,z)$ using 4 metrics; Chamfer distance (CD), sliced Wasserstein distance ($SW_1$)[64], and differences in total surface area ($\Delta A$) and volume ($\Delta V$) (Extended Fig. 3d, Methods). Considering inevitable rasterization errors when mapping the floating-point precision 3D sphere $S_\Omega^2(x,y,z)$ to $S(u,v)$ defined on an integer $(u,v)$ image grid, we measured low vertex position errors according to CD and $SW_1$. Cells with lamellipodia had the lowest error (median CD=1.77 voxel, $SW_1$=0.93 voxel) and cells with blebs were slightly worse (median CD=2.79 voxel, $SW_1$=4.28 voxel), likely due to their intrinsically small height (small topographic $d$). As one would expect, cells with long, thin filopodia displayed the largest discrepancies (median CD=10.33 voxel, $SW_1$ =18.45 voxel). Correspondingly we measured a small $\Delta A$ (+1.2%) and $\Delta V$ (+7.9%) for cells with lamellipodia. $\Delta A$ was larger for cells with blebs (-11.6%) and measured to be too large for filopodia (-55.3%) when compared to $\Delta V$ differences measured after making $S(x,y,z)$ watertight (+4.2% blebs, +3.4% filopodia). Visualization of exemplar cells show good geometric correspondence between $S_{\text{topo}}(x,y,z)$ and $S(x,y,z)$ (Extended Fig. 3e). Salient surface features were largely captured, albeit smoothened and blurred in $S_{\text{topo}}(x,y,z)$ when local surface regions were underrepresented due to being distant relative to $S_{\text{ref}}(x,y,z)$ (Extended Fig. 3e, black triangles, 1st row blebs and 4th row lamellipodia). Most of the primary morphological features, namely the length and thickness of long, thin filopodia (Extended Fig. 3e, green triangles), except those located both densely together and distant relative to $S_{\text{ref}}(x,y,z)$ (Extended Fig. 3e, red triangles), were captured. u-Unwrap3D was able to capture both the complex lamellipodia folds and curved cell bodies to high accuracy (Extended Fig. 3e, row 2,4). Closer inspection of $S_{\text{topo}}(x,y,z)$ and $S(x,y,z)$ in these cells traced a large $\Delta A$ to meshing errors in the input surface $S(x,y,z)$, which caused internal volumetric structures to be merged into the cell surface representation, and overestimation of total surface area. These errors affect the CD and $SW_1$ to lesser extent.

In summary, our results demonstrate that u-Unwrap3D is robust and applicable to process unconstrained geometries. For maximum resolution of high curvature surface features, a genus-0 reference surface $S_{\text{ref}}(x,y,z)$ proximal to the input surface $S(x,y,z)$ is recommended with a large $(u,v)$ grid size $N$ and small $\alpha$ step sizes when propagating $S_{\text{ref}}(u,v)$. However, these choices depend on the quality of the input surface mesh $S(x,y,z)$, which depends on the robustness of cell segmentation in the face of noisy image raw data.

**u-Unwrap3D enables unsupervised instance segmentation of subcellular surface motifs**
The unbiased identification and segmentation of individual protrusive features in unconstrained 3D surface geometries is nontrivial. Cellular protrusions present complex morphological characteristics that are difficult to define descriptively. Even well-known morphological motifs exhibit significant heterogeneity and ambiguity. Not all blebs are spherical, lamellipodia are often plate-like with high curvature ridges but otherwise have no readily-defined shape prior, and filopodia, though long and thin, can sprout haphazardly from elevated 'stumps' or even off of each other. In areas of dynamic and dense protrusions, where does one protrusion start and another end? Consequently, most existing approaches focus on particular surface features of interest such as 'ridge' networks that can be segmented by designed imaging filters or through trained semantic segmentation, with morphological processing and parameter tuning[46,47,65,66]. With u-Shape3D we introduced a multi-class morphological motif detection by partitioning the 3D surface into convex patches and applying support vector machines trained with expert annotation to classify the patches into pre-specified motif types[45]. However, this approach cannot detect and segment all protrusions generally, only the limited motifs for which the supervised classifier has been trained on. Lastly, even after obtaining the segmentations, how do we systematically measure salient protrusion properties? For example, with respect to what reference surface should protrusion height be measured? Where is the protrusion width to be measured? How is the internal volume of a protrusion determined?

These segmentation and characterization problems can be significantly better defined in the topographic 3D surface representation $S(d, u, v)$, which captures in one field-of-view all surface features protruding normally to $S_{\text{ref}}(x, y, z)$. As $d$ preserves the total Cartesian 3D curvilinear distance from $S(x, y, z)$ to $S_{\text{ref}}(x, y, z)$ along the gradient of steepest descent we can formally define a 'protrusive' feature as having a $d$-coordinate greater than that of a reference topographic surface, $S_{\text{ref}}(d_{\text{ref}}, u, v)$, and measure the protrusion height as the difference, $h = d - d_{\text{ref}}$. For example, protrusive features could be specified as those having $h > \bar{h}^{cMCF}$, the mean height of $S(d, u, v)$ relative to the planar topographic 3D cMCF surface $S_{\text{ref}}^{\text{cMCF}}(d, u, v)$ (i.e. $d_{\text{ref}} = f(u, v) = 0$). However, this definition leads to under-segmentation (Extended Fig. 4a). A remedy would be an intermediate surface $S_{\text{ref}}(d_{\text{ref}} = f_{\text{smooth}}(u, v), u, v)$, which interpolates between the input rugged topographic cell surface, $S(d, u, v)$ and the 2D planar cMCF cell surface $S_{\text{ref}}^{\text{cMCF}}(d, u, v)$, (Fig. 3a,b). Whereas this problem is difficult to frame in Cartesian 3D, in the topographic space the interpolation can be solved naturally by using asymmetric least squares (ALS) optimization with a Whittaker smoother[67,68], where the asymmetric weights allow us to account for the heterogeneous protrusion height; and the desired level of surface smoothness can be incorporated as a regularization term (Extended Fig. 4b, Methods). To use ALS, we create a $(u, v)$-parameterized approximation of $S(d, u, v)$ with $d \simeq f(u, v)$ using straightforward image processing procedures (Extended Fig. 4b, Methods). By exploiting these properties of the topography space, we developed a general approach to segment any protrusion motif including blebs, lamellipodia and filopodia (Extended Fig. 4c,d) with minimal heuristic parameters to tune. Importantly, we did not need to design specialized image filters[65,66], compute and cluster feature descriptors[46,69-71], or require data training[45-47].

We demonstrate the segmentation of individual protrusion instances, capturing motifs identified by uShape3D, but without the need for training annotations. By construction, in topographic 3D $(d, u, v)$ space all surface protrusions are oriented upwards with increasing $d$. Moreover, the tops of protrusions are individually separated as local regions of high topographic mean curvature, which we identify by thresholding and applying connected component labelling in the topographic volume, $V(d, u, v)$ (Fig. 3c). Mapping the segmented regions back onto the topographic mesh $S(d, u, v)$, we diffuse these initial 'seed' labels across the surface using a combined geodesic distance and dihedral angle affinity matrix to naturally segment the 'stem' of the individual protrusions (Methods). The dihedral angle measures the discontinuity in local mean curvature. It incorporates the prior intuition that the boundaries of a label should expand faster on local surfaces of homogeneous curvature such as that on a 'hill', compared to another label experiencing large curvature differences in its local surface region, such as in a valley between multiple 'hills'. The combined affinity matrix thus introduces a morphology-aware competition between segmentation labels and provides a biophysical rationale for defining which surface patches belong to individual 'seed' protrusions. Furthermore, the dihedral angle is large between a protrusion and the main cortical cell surface and thus serves as a soft stopping criteria for diffusion (Extended Fig. 4d) in addition to applying the binary protrusion segmentation from above. Lastly, we filter out protrusions that are too small and close any small holes using the Cartesian 3D surface area. The final segmentation result qualitatively and quantitatively agrees with that obtained by supervised u-Shape3D for lamellipodia (Fig. 3d) but yields more contiguous labels and is less prone to over-segmentation (Fig. 3e). Importantly, this segmentation strategy is applicable, even when not all protrusions are equiareally represented in $S(d, u, v)$, as shown by the segmentation of the majority of blebs and filopodia in exemplar cells (Extended Fig. 4d).

Individual segmented protrusions are genus-0 open-surface 3D submeshes that can be directly mapped to the 2D plane (Fig. 3f). This allows us to further refine the segmentation, for example, by detecting and splitting under-segmented blebs by a gradient watershed algorithm (Methods). Thanks to bijectivity, the refined segmentation labels can be transferred back onto the original surface mesh (Fig. 3f,g, c.f. before and after refine, black triangles). Both the before (adjusted normalized mutual information, NMI=0.57) and the after refinement (adjusted NMI=0.54) segmentations agree with u-Shape3D. However, u-Unwrap3D segmented blebs are more complete, with more blebs of larger surface area (150 blebs in total). In contrast, u-Shape3D over-segments small blebs (742 blebs) that were found to originate from erroneous meshing of internal structures in $S(x, y, z)$ (see Extended Fig.3e). This example illustrates the potential pitfalls of identifying motifs from local surface patches only with potentially imperfect input 3D meshes. In contrast, with u-Unwrap3D any surfaces internal to the cell volume are readily removed when mapped into topography as these surfaces have $d$-coordinates less than the reference surface, $S_{\text{ref}}(d_{\text{ref}} = f_{\text{smooth}}(u,v), u, v)$.

Finally, the representation $S(d, u, v)$ enables partitioning of the input cell volume into the sum of a reference volume representing the underlying cortical cell body and the unique volume occupied by individual protrusions (Fig. 3h). We do so by $(u, v)$-parameterizing the reference cortical surface submesh, $S_{\text{ref}}(d, u, v)$ after removing all individual protrusion submeshes as a grayscale image such that the pixel intensity value at $(u, v)$ is the respective $d$-coordinate, and inpainting the missing $d$-coordinates at $(u, v)$ coordinates corresponding to the subtracted protrusions to generate a full reference binary volume (Extended Fig. 4e, Methods). For the protrusions, we first devised a marker-controlled lateral watershed depth propagation to diffuse the surface-based protrusion segmentation labels uniquely throughout the full topographic $(d, u, v)$ space slice-by-slice, top-to-bottom (Extended Fig. 4f). Then, individual protrusion volumes were generated by masking the propagated label volume with the reference binary volume (Extended Fig. 4g). We compared our topography-guided decomposition strategy to a fully Cartesian 3D mesh processing approach whereby individual protrusion submeshes were first closed by constructing a surface patch that minimized the local bending (or harmonic) energy[72] (Methods). The closed reference volume was then generated using all such patches to impute the residual holes in the raw reference surface. Fig. 3i panels i-ii show that the computed surface area (slope=0.90) and volume (slope=0.95) of individual protrusions are similar for both methods. However, the 3D mesh processing protrusions consistently under-measure larger protrusions. Crucially, the imputed reference volume appears artefactual. Where protrusions were located, the surface is overly smooth, and even involuted. These regions contrast starkly with non-imputed surface areas between protrusions, creating artificial 'peaks' and 'ridges' of high mean curvature (black arrows). In comparison, the topography-guided reference volume is mechanically more plausible.

In summary, the ability to map freely between topographic and Cartesian 3D surfaces and their respective volumetric representations enabled us to design simple and generalizable methods to detect and segment in an unsupervised fashion individual morphological motifs from unconstrained surface geometries. To bijectively map the topographically segmented surface protrusion labels onto the 2D plane, which is an optimal representation for tracking the segmented motifs, we developed a topographic cMCF for u-Unwrap3D (Fig. 3j, Suppl. Video 6, Methods).

**u-Unwrap3D enables tracking of 3D subcellular surface motifs and molecular activity in 2D**

A central goal of live cell imaging in 3D is to visualize the spatiotemporal relations between molecular activities and cell behaviors, including morphodynamic outputs. Progress has been made on software developments that allow unbiased and statistically meaningful analysis of cell morphology and molecular distributions[45,47,60,73,74]. However, to remain algorithmically and computationally tractable these pipelines have been restricted to quasi-static representations of dynamic processes. Surface-guided computing with u-Unwrap3D allows us now to remedy this limitation. Dynamic behaviours on complex 3D cell shapes, including their morphological and molecular signal activity changes, can be mapped to 2D representations where powerful analytical pipelines exist for spatiotemporally consistent tracking. Results can be statistically evaluated and, if of interest, be visualized in 3D by leveraging the bijective properties of u-Unwrap3D mappings. Fig. 4 and 5 demonstrate this capacity based on two examples.

**Individual blebs dynamically recruit Septins to local surface regions during retraction**

We first analysed potential relations between dynamic surface blebbing and the recruitment of Septins. Blebs are globular membrane protrusions of 1-2 μm diameter that are thought to extend in areas of localized membrane detachment from the actin cortex[75,76]. Intracellular pressure expands the budding blebs outward, followed by rapid assembly of a contractile actomyosin network that yields retraction. Cycles of protrusion and retraction have been described to last a few tens of seconds. Associated with the cycles are molecular activities both driving and responding to the morphological dynamics. One such process is the assembly of Septin protein polymers at sites of negative curvature (from a cell-external perspective) emerging at the bleb necks. Our previous work[38] has shown that disrupting the bleb cycle diminishes Septin assembly at the cell surface. Here we now exploit the ability of u-Unwrap3D to track individual bleb cycles and quantify Septin accumulation by remapping surface morphology and a fluorescent marker of Septins to an appropriate 2D representation. We acquired 3D volumes of SEPT6-GFP-expressing MV3 melanoma cells every 1.2s for 200 timepoints. As the cortical cell body exhibits little temporal variation and blebs protrude normally to the surface, the temporal mean cell surface, $\bar{S}(x, y, z)$ is a good proxy of the cell cortex. We apply u-Unwrap3D to $\bar{S}(x, y, z)$ to create a common static $(d, u, v)$ coordinate space for computing topographic 3D $S(d, u, v)$ and 2D planar $S(u, v)$ representations for each timepoint. $\bar{S}(x, y, z)$ was computed by meshing the mean binary volume across individual binary voxelizations of the cell surface over all 200 timepoints (Methods). Note the construction of a common topographic space from a single mesh for a timelapse is computationally efficient but applicable only if cell shape changes lie within the Cartesian subvolume mapped by the topographic space. For large shape changes u-Unwrap3D should be applied to individual timepoints and spatiotemporal registration used to align surfaces to a common reference using $S_{\text{ref}}(x, y, z)$ or $S(u, v)$ representations. The segmentation tools discussed above were used to detect all bleb instances from $S(d, u, v)$ and mapped to $S(x, y, z)$ and $S(u, v)$ for each timepoint (Fig. 4a, Suppl. Video 7). Similarly, the computed mean curvature and the normalized Septin intensity surface signals $F_i(S(x, y, z))$ from Cartesian 3D were mapped to topographic 3D $F_i(S(d, u, v))$ and into 2D, $F_i(S(u, v))$, to enable simple bleb tracking and timeseries analysis.

To track blebs in 2D we computed the bounding box of individual bleb instances in every timepoint after appropriate image padding to account for spherical periodicity (Fig. 4b left). In $S(u, v)$ bleb dynamics can readily be followed by an established 2D multi-object bounding box tracker[77] with mean curvature optical flow-guided bipartite matching (Fig. 4b middle, Methods). The bijective mapping allows us to map the resulting trajectories from $(u, v)$ coordinates (individually colored) to $(x, y, z)$ coordinates to generate bleb tracks in Cartesian 3D (Fig. 4b right, Suppl. Video 7). Due to the fast temporal acquisition rate, only a northern portion of the cell can be maintained in-focus. Again, u-Unwrap3D's bijectivity between 3D and 2D representation enabled us to easily map a manually annotated out-of-focus subvolume onto $S(x, y, z)$ and into $S(u, v)$ to retain for analysis only the bleb tracks that remain within the in-focus surface regions (Fig. 4c). The distribution of individual bleb track lifetimes showed a peak at 14s and a long tail up to 240s, suggestive of a mixture of short- and long-lived blebs (Fig. 4d). The mean bleb lifetime of 27s corresponded well with a 30s periodicity given by the first peak of the temporal autocorrelation of mean curvature $H(S(x, y, z, t))$ in Cartesian 3D (Fig. 4e). This validates at the population level the accuracy of single bleb tracking after projection and segmentation of mean surface curvature in 2D.

The temporal autocorrelation curves of mean curvature and Septin intensity showed a high level of similarity, suggesting co-fluctuation of the two surface signals. Indeed, we had previously shown that surface regions of high Septin intensity with negative surface curvature for at least 30s display a correlation between negative curvature value and Septin intensity[38]. Whilst the majority of Septin pulses endured only one cycle of bleb formation and retraction, *de novo* formation of stable Septin structures appeared to be driven by several Septin pulses occurring in short succession. We thus hypothesized that these were formed by iterative bleb-driven curvature generation events resulting in local levels of Septin oligomers surpassing a threshold necessary for inter-oligomer polymerization and enabling stabilization through formation of higher-order structures[38]. The ability to spatiotemporally track individual blebs enabled us now to quantitatively test this model. For the duration of each tracked bleb, we sampled within the 2D bounding box distortion-corrected timeseries of bleb surface area, on-/off- bleb surface mean curvature and Septin intensity (Methods). We used the 2D $S(u, v)$ bounding box to define a bleb's spatial area-of-influence and its surface area. The Cartesian 3D $S(x, y, z)$ box area was taken as the bleb's 3D surface area in each tracked frame. Within the 2D bleb bounding box, 'on-bleb' is the largest spatially contiguous region of high mean curvature. The remainder area is 'off-bleb'. A single curvature threshold was computed by 3-class Otsu thresholding over all $H(S(u, v, t))$ to define the regions of low/high mean curvature in $S(u, v, t)$. Using the extracted timeseries we reconstructed the temporal profile of bleb area, on-/off- mean curvature and Septin intensity of a single mean bleb event in a window of 35s centered on the timepoint of maximum bleb area averaged over 545 single bleb events from 480 bleb tracks. We sample $\pm 17.5$s before and after the timepoint of maximum bleb area to exceed the 30s periodicity inferred from temporal autocorrelation by 5s and capture the full dynamics. We then broke the mean timeseries into four distinct temporal phases of bleb-mediated Septin recruitment, each ≈5s long (Fig.4f labels A-D). In phase A, the bleb begins to expand, increasing its surface area, accompanied by a sharp increase in on-bleb $H$ and a decrease in Septin intensity as the plasma membrane detaches from the actin cortex. The expansion also reduces off-bleb $H$, causing a decrease in Septin intensity off-bleb, presumably due to disrupting Septin structures. In phase B, the bleb reaches maximum size and then begins to retract with decreasing surface area. Interestingly, unlike mean curvature, which begins to decrease before the maximum bleb area, the change in area is symmetrical, occurring $\pm 2.5$s relative to the time of maximum bleb area. Coincident with the bleb increasing to a maximum area, off-bleb $H$ decreases to a minimum and

Septin intensity both on/off bleb stabilizes at a minimum. As the bleb retracts and the actin cytoskeleton reassembles, off-bleb $H$ increases and Septin intensity begins to increase both on/off bleb. In phase C (+2.5s to +8.0s after peak bleb area), the bleb area continues to decrease but at a slower rate. Unexpectedly, the off-bleb negative curvature $H$ plateaus at a value lower than its starting value (before phase A) instead of continuing to increase. Concurrently, Septin intensity undergoes the greatest rate of increase on both on-bleb and off-bleb surfaces such that at the end of phase C, the Septin intensity is at the levels before phase A (c.f. -15s to -10s). In phase D (+8.0s to +14.0s), bleb area and on/off-bleb $H$ recover to pre-expansion levels while Septin intensity continues to increase before plateauing on both on/off-bleb surfaces. Beyond phase D, the next bleb cycle begins, with similar temporal characteristics to phase A. Altogether these results indicate that blebs generate optimal curvature dynamics during retraction to recruit Septin polymers to the surface regions around blebs. Moreover, the data support our model of Septins being recruited to negative curvature patches in a cyclic fashion, where each bleb formation-retraction drives the local accumulation of a few more oligomers until a threshold concentration is reached to trigger inter-oligomer polymerization of a stable Septin assembly. Notably, none of these observations could have been made without u-Unwrap3D.

**Ruffles are driven by locally enriched filamentous actin and migrate actively on the cell surface**

We also applied the capacity of u-Unwrap3D to examine putative relations between the dynamics of membrane ruffles, filamentous actin and surface actin retrograde flow. Membrane ruffles are thin, rapidly-moving, actin-rich protrusions[78,79]. They are thought to play a role in cell migration, and it has been proposed that ruffles arise as a consequence of inefficient adhesion in cellular lamellipodia[80]. Yet, it is unclear whether they have a specific function in migration or elsewhere, and the precise molecular and mechanical mechanisms of ruffle formation remain poorly understood. Although membrane ruffles have been the showcase for many of the recent advances in volumetric light sheet microscopy, there are no tools for quantitative assessment of membrane ruffling[78,81]. The thin, lamellar appearance make ruffles extremely challenging to segment in 3D live-cell images[46]. Moreover, ruffles are transient and they exhibit significant heterogeneity in nature and distribution across the cell surface. These characteristics make ruffles difficult if not impossible to directly track in Cartesian 3D.

There are many unanswered questions about cell membrane ruffles. Here, we focus on defining a membrane ruffle by a set of objective criteria and then on measuring their speed. We acquired 3D volumes of SU.86.86 pancreatic ductal adenocarcinoma cells plated on fibronectin-coated cover glasses and co-expressing Tractin-mEmerald and myristoylated CyOFP1. Tractin is a marker for actin[82], while myristoylated CyOFP1 served as a diffuse cell membrane marker. We acquired images every 2.27s for 30 timepoints, and used u-Unwrap3D to map the Cartesian 3D $S(x,y,z)$ surface of every timepoint into topographic 3D, $S(d,u,v)$ and into 2D, $S(u,v)$ for tracking (Fig. 5a, Suppl. Video 8). The cortical shape change was relatively small but ruffles travel on the surface. Like for bleb tracking we constructed a common static topographic coordinate space $(d,u,v)$ for all time points. Here, we used the reference surface, $S_{\text{ref}}(x,y,z,t=0)$ for this purpose (Extended Fig.5a). We then sought to track the spatial location of ruffles as individual 'ridge' objects. To do so, we projected $S(d,u,v)$ onto the 2D plane, i.e. the $(u,v)$ coordinate was the same as setting $d=0$, $(d=0,u,v)$ instead of applying topographic cMCF. We computed the local enrichment of the actin signal as the ratio of Tractin-mEmerald to CyOFP1 intensity. Kymograph visualization of $F_i(S(u,v,t))$ for mean curvature, $H$ and Tractin-mEmerald/CyOFP1 (TC) showed strong co-fluctuation of curvature and local actin intensity. It also highlighted the transient, 'ripple-like' nature of ruffles and their merging and dissipation with a delay between successive ruffles of ≈20s (Fig. 5b). To avoid complex image processing operations such as merging and splitting we applied optical flow-based region-of-interest (ROI) tracking[83,84] to the TC intensity to track simultaneously the protrusive membrane ruffling and retrograde surface actin flow (Fig. 5c, left). The $(u,v)$ image grid size was 1025 x 512 pixels (not 1024 x 512 as ruffles required active contour cMCF to be used to generate $S(d,u,v)$, see Methods) and the average dimensions of the tracked $(u,v)$ ROI size was 23 x 23 pixels (approximately 0.06 x 0.11 µm pixel size in physical space). The resulting 2D ROI tracks exhibit unidirectional motion towards the cell center as shown by coloring directionality and remapping of $(u,v)$ ROI tracks to polar $(r,\phi)$ (Fig. 5c, middle, Extended Fig. 5b) and Cartesian 3D $(x,y,z)$ coordinates (Fig. 5c, right) (Suppl. Video 8). The optical flow tracks measure the geodesic cellular surface speed. To measure specifically the component corresponding to mean lateral ruffle travel speed we must project the 3D optical flow velocities along the same plane as the flat cell bottom. The volumetric imaging of cells is acquired on a cover glass tilted at ≈45⁰ (Extended Fig. 5c, left). Direct 3D plane-fitting to $S(x,y,z)$ to determine a precise angle is sensitive to outlier points and shapes deviating from an elongated ellipsoid. Thanks to the $S(u,v)$ representation of u-Unwrap3D mapping the curved proximal cell surface and the planar cell bottom to the upper and lower half of the unwrapped 2D image respectively, we could readily annotate in 2D $(u, v)$ the cell bottom and fit a 3D plane to only this surface patch in $(x, y, z)$ (Extended Fig. 5c, middle, right). Similarly, we could gate ROI tracks in $(u, v)$ and compute the lateral speed only for those associated the lamellipodia and lamella surface. Doing so we found two populations in the speed histogram (Fig. 5d). Visualizing the speed on $S(x, y, z)$, the faster population corresponds to higher curvature ruffles with speeds ranging from 2-10 µm/min, and an average speed of 4.2 µm/min (Suppl. Video 8). This is at least two times faster than the slower population surface retrograde actin flow ranging from 0-1 µm/min, which are consistent with the flow speeds we used to measure by 2D fluorescent speckle microscopy in the lamella of epithelial cells[85]. This result suggests ruffles are actively produced and transported across the cell surface. To assess the synchronicity of actin and ruffles we extracted distortion-corrected timeseries of TC and mean curvature, $H$ by sampling and averaging the respective values within a spatial window of 23 x 23 centered around the $(u, v)$ coordinates of all ROI tracks on the lamellipodia and lamella surface. Averaging the temporal cross-correlation curves (mean±95% confidence interval) of individual ROI tracks we find a significant positive instantaneous (lag=0) correlation of 0.2 (Fig. 5e, left). Plotting the instantaneous (lag=0) correlation values of individual ROI tracks as a function of the mean $H$ value of the same ROI track and visualizing the instantaneous correlation on $S(x, y, z)$, we found that ruffles with higher positive surface curvature are more temporally correlated with TC intensity (Fig. 5e, middle, right). Altogether our results show that ruffles are highly dynamic, transient protrusions that actively migrate on the cell surface and driven by locally enriched filamentous actin. Our u-Unwrap3D framework provides now the platform for systematic investigation of the mechanisms that drive and regulate these dynamics.

**Discussion**

Analyzing the spatiotemporal organization of molecular distributions and signaling activities on cell surfaces in 3D has been limited by the lack of methods to represent, track and process these dynamics. Here we introduce a surface-guided computing framework, referred to as u-Unwrap3D, to bijectively map a genus-X Cartesian 3D surface to equivalent surface and volume representations that are optimally suited for a distinct analytical task. The mappings rely on two critical insights: i) the engineered surface deformation of $S(x, y, z)$ to generate a genus-0 $S_{\text{ref}}(x, y, z)$ for which a 3D spherical parameterization exists; ii) a novel, efficient algorithm to relax geometric distortion on the 3D sphere in a bijective and tunable manner. Insight i) is fundamental to allowing $S_{\text{ref}}(x, y, z)$ to serve as a representative proxy that captures all the salient surface features of the genus-X $S(x, y, z)$ surface and insight ii) to preserving this property when $S_{\text{ref}}(x, y, z)$ is unwrapped to the 2D plane. Using diverse cell examples we validated that u-Unwrap3D could be widely applied (>90% of cases) in a manner robust to the input surface mesh quality and that it accurately captures the cell geometry to transfer salient surface features, morphological or molecular, between all representations. We note that this 90% is an underestimation of the applicability. The validation dataset was assembled to be deliberately heterogeneous and was not segmented with the downstream aim of surface mapping. In practice, segmentation algorithms continue to improve and postprocessing techniques can be used to create improved surface meshes. Moreover we can leverage surface meshing algorithms that are more sophisticated than marching cubes, such as dual contouring[86] and shrink-wrapping[87], to guarantee watertight mesh creation. For timelapse acquisitions, as we showed in Fig. 4,5, the situation is even simpler. Only one timepoint or average surface is needed to generate a single common $\bar{S}_{\text{ref}}(x, y, z)$ to unwrap all timepoints. u-Unwrap3D puts in place a generic platform for the spatiotemporal processing of unconstrained cell geometries.

u-Unwrap3D is applicable for arbitrary genus-X 3D surfaces, $S(x, y, z)$ wherever a genus-0 reference surface, $S_{\text{ref}}(x, y, z)$ can be found from conformalized curvature flow (cMCF). It works best when $S_{\text{ref}}(x, y, z)$ is close to the $S(x, y, z)$. In our current implementation, the generation of $S_{\text{ref}}(x, y, z)$ may be suboptimal or fail when either small handles in the mesh pinch together, causing early termination of the cMCF, or if the binary voxelization and morphological hole closing fails to infill large handles and holes in the cMCF $S_{\text{ref}}(x, y, z)$. Whilst we can increase the range of morphological hole closing in the latter case, we restrict dilation to a maximum 3-5 voxels to minimise smoothing out protrusion features in $S_{\text{ref}}(x, y, z)$. Mesh surgery methods have been developed in computer graphics to make non-watertight 3D meshes watertight[88,89], and fix imperfections like holes and handles to reduce genus[90,91] to generate higher quality 3D meshes. These have yet to be applied and fully tested for complex cell surfaces. Future development will investigate how to apply such procedures to allow u-Unwrap3D to be applied to input meshes of any quality. In our data we have paid attention during the acquisition to generating sufficient foreground-to-background contrast for reliable surface segmentation, thus minimizing mesh defects.

The concept of geometrical reduction of 3D into 2D geometry through the choice of an optimal coordinate transformation has long existed in mathematics and physics to simplify mathematical manipulation and plotting. For example, parametric coordinates describe the sphere, the cylinder, Mobius strip and helicoids amongst others[92]. In computer graphics this is realised in the common practice of mapping between surfaces through simpler intermediaries: for example, the texture mapping of arbitrary surfaces by optimal surface cutting and mapping of the cuts into individual 2D shapes[57] and $(u, v)$ surface parameterization by cutting and gluing individually mapped 2D planar patches[93] or mapping to canonical shapes such as the triangle[94], plane[39] or polyhedra[95] to minimize distortion. u-Unwrap3D draws inspiration from this thinking. Through the availability and development of rationalized multiple 3D-to-2D representations, u-Unwrap3D projects analyses that would otherwise require specialised mathematical operations into a sequence of simpler, computationally tractable procedures for 3D mesh processing, image processing and machine learning with specific consideration for single cell biology. Unlike computer graphics benchmarks, surface protrusions are irregular, high curvature and dense. First, we specifically chose representations that map the whole cell surface with well-behaved topologies such as the sphere and plane and designed a relaxation scheme to guarantee interpolation between the minimal conformal and area distortion. This bypasses the numerical instabilities of stitching multiple surface maps and only quasi-conformal mappings for the majority of literature methods. Second, although $S_{\text{ref}}(x, y, z)$ was conceived as a mathematical trick to enable spherical parameterization, because it is derived explicitly from $S(x, y, z)$ and not as a canonical shape, it serves biologically to decompose $S(x, y, z)$ as the 'sum' of a smooth cell cortex and surface protrusions. Our results suggest the introduced $S_{\text{ref}}(x, y, z)$ and $S_{\text{topo}}(x, y, z)$ could open up new opportunities to model and quantify the interplay of dynamic membrane morphology and associated signals with more volumetric nuclear/cytoplasmic signalling. Lastly, we designed mappings to underlying representations, including Cartesian 3D, 3D sphere and 2D plane that are standard inputs in computer vision and machine learning.

With u-Unwrap3D standard computer vision, machine learning methods become directly applicable to computing tasks on rugged complex surfaces. More recently, research into combining different geometric representations are state-of-the-art in addressing complex computational problems such as multiple 2D image views to inform 3D mesh reconstruction[96,97], or 3D mesh vertex coordinates with 2D unwrapped images for feature extraction[98,99]. u-Unwrap3D is fully complementary to these research developments - unifying the different representations into a single surface-guided computing framework for downstream analysis. The u-Unwrap3D framework is made available as a Python library. The resources and validation provided by this work will aid the cell biology community to generate testable hypotheses of the spatiotemporal organization and regulation of subcellular geometry and molecular activity.

**Acknowledgments**
We thank Hanieh Mazloom Farsibaf and Zach Marin for discussions on mesh processing. We also thank Qiongjing Zou for putting the Python library into a continuous integration coding framework. Funding for this work in the Danuser lab was provided by the grant R35 GM136428 (NIH). AW is a fellow of the Jane Coffin Childs Memorial Fund. GMG is a Damon Runyon Fellow (DRG 2422-21).

**Author Contributions**
FYZ conceived and developed u-Unwrap3D and conducted the analyses. MD generated the uShape3D protrusion instance segmentations and surfaces used in validation. AW acquired timelapse of the blebbing MV3 cell. GMG, BJC and BC acquired timelapse of the ruffling SU.86.86 cell. FYZ and GD wrote the manuscript with input from all authors. GD provided supervision and obtained funding.

**Competing Interests**
The authors declare no competing interests.

**Data Availability**
All data presented herein are available from the corresponding author upon request.

**Code Availability**
u-Unwrap3D is open source, developed as a Python library and is available at: https://github.com/DanuserLab/u-unwrap3D.

## Methods

### u-Unwrap3D framework

Following we describe the algorithms underpinning each step of u-Unwrap3D depicted in Fig.1.

Step 1: $S(x, y, z)$ to $S_{\text{ref}}(x, y, z)$

*Conformalized mean curvature flow (cMCF).* cMCF[49] modifies the mean curvature flow (MCF) to avoid the formation of pinches and collapsed vertices that compromise bijectivity and cause early flow termination for watertight meshes with high curvature features. We find cMCF also reduces the size of holes and handles in $S(x, y, z)$. MCF evolves $\Phi_t$, the mesh at time $t$ according to $\frac{\partial \Phi_t}{\partial t} = \Delta \Phi_t$, where $\Delta$ is the Laplace-Beltrami operator induced by the metric $g_t$ and $\Phi_t(p) = \sum_{i=1}^{N} v_i(t) B_i(p)$ is the discrete mesh parameterization with $N$ vertex positions $v_i(t) = \{v_1(t), \dots, v_N(t)\} \subset \mathbb{R}^3$, each $v_i(t)$ a 3D $(x_i(t), y_i(t), z_i(t))$ coordinate tuple and $\{B_1, \dots, B_N\}$ the local function basis, which for a triangle mesh is the linear hat basis spanned by the edge vectors. Galerkin's method[100] is used to find a weak, least-squares solution to the MCF equation within the span of $\{B_i\}$, by solving $\int_S \left( \frac{\partial \Phi_t}{\partial t} \cdot B_i \right) d\mu_t = \int_S (\Delta_t \Phi_t \cdot B_i) d\mu_t$, $\forall\, 1 \leq i \leq N$. $S$ is the surface spanned by $\{B_i\}$ and $d\mu_t$ the volume form. The equation is solved to obtain the vertex position, $v(t + \delta t)$ at the next iteration, $t + \delta t$ using backwards Euler integration; $\frac{\partial \Phi_t}{\partial t} \approx (\Phi_{t+1} - \Phi_t)/\delta_t$, $\Delta \Phi_t \approx \Delta \Phi_{t+1}$ and noting the Laplace-Beltrami is the divergence of the gradient, $\Delta = \nabla \cdot \nabla$ with respect to the local mesh metric $g_t$ to get $\int_S \left( \sum_{i=1}^{N} v_i(t + \delta t) B_i \cdot B_j - \sum_{i=1}^{N} v_i(t) B_i \cdot B_j \right) d\mu_t = \delta_t \int_S \sum_{i=1}^{N} v_i(t + \delta t) g_t(\nabla_t B_i, \nabla_t B_j) d\mu_t$. The integrals, $M_{ij}^t = \int_S B_i \cdot B_j \, d\mu_t$ and $L_{ij}^t = \int_S g_t(\nabla_t B_i, \nabla_t B_j) d\mu_t$ are called the mass ($\mathbf{M}_t$) and Laplacian ($\mathbf{L}_t$) matrices at time, $t$. Substituting this notation and rearranging, the linear algebra MCF equation is

$$MCF \coloneqq (\mathbf{M}_t - \delta_t \mathbf{L}_t) v(t + \delta t) = \mathbf{M}_t v(t)$$

$v(t + \delta t)$ is then computed from $v(t)$ by direct matrix inversion. cMCF modifies the MCF equation above by using the Laplacian matrix at time $t = 0$, $\mathbf{L}_0$ for all timepoints. The Laplacian matrix is a measure of stiffness between local mesh faces, see active contour cMCF below. Using $\mathbf{L}_0$ for all timepoints instead of recomputing implicitly constrains mesh faces to retain the same aspect ratio and this conformalizes the flow.

$$cMCF \coloneqq (\mathbf{M}_t - \delta_t \mathbf{L}_0) v(t + \delta t) = \mathbf{M}_t v(t)$$

We use the *libigl* library[101] with the cotangent Laplacian and barycentric mass matrix as the default implementations of $\mathbf{M}_t$, $\mathbf{L}_t$ respectively. We improve the numerics of solving cMCF by normalization of the surface area and recentering of vertex coordinates at the origin after each iteration as recommended in Alec Jacobson's blog post (https://www.alecjacobson.com/weblog/?tag=mean-curvature-flow). u-Unwrap3D implements for optional usage the robust Laplacian of Sharp et al.[102] instead of the cotangent Laplacian, which can improve numerical stability.

*Automatic stopping criterion for cMCF.* The ideal reference surface $S_{\text{ref}}(x, y, z)$ for topographic representation is the cortical cell shape without protrusions. We find that this corresponds to finding the 'elbow point' in the mean absolute Gaussian curvature (Fig. 1a) and not the convergence limit of cMCF which is the sphere[49]. The difference in the mean absolute Gaussian curvature over vertices, $\Delta \overline{|K_t|} = \overline{|K_t|} - \overline{|K_{t-1}|}$ between successive iterations $t - 1$ and $t$ is used as an automatic stopping criterion for cMCF, $\text{stop}_t = \max(t_{\min}, t_K)$. $t_{\min}$ is a user-specified minimum iteration number and $t_K$ is the first iteration for which $\Delta \overline{|K_t|}$ exceeds a user-specified threshold, $\Delta \overline{|K_t|} > \Delta_{thresh}$. We compute the discrete Gaussian curvature, $K_{v_i}$ at a vertex $v_i$, given by the vertex's angular deficit[103], $K_{v_i} = 2\pi - \sum_{j \in N(i)} \theta_{ij}$ where $N(i)$ are the triangles incident on $v_i$ and $\theta_{ij}$ is the angle at vertex $i$ in triangle $j$.

*Mesh voxelization and remeshing.* Mesh voxelization converts a surface mesh $S(x,y,z)$ to a binary volume image $\beta$ where individual voxels are either 1 if they are interior to the surface or 0 if exterior. To do this we create a $X \times Y \times Z$ voxel volume image larger than the surface with voxels initialised to 0. We then set the intensities of all voxels indexed by the mesh $(x,y,z)$ barycenters to 1, i.e. $B(x,y,z) = 1$. To ensure a closed binary volume with all interior voxel intensities = 1, $S(x,y,z)$ was iteratively subdivided by replacing each triangle face by four new faces formed from adding new vertices at the midpoint of every edge until the mean triangle edge length is < 1 voxel. We use the barycenter coordinates of the final mesh $S_{\text{final}}(x,y,z)$ to set the binary voxel values. In case of small holes in $S(x,y,z)$ that would prevent a closed binary volume by binary filling only, $\beta$ was first dilated with a ball kernel, then binary infilled, and lastly binary eroded with a ball kernel. The postprocessed $B$ was meshed using marching cubes[104] followed by construction of an approximated centroidal voronoi diagram (ACVD)[105] to produce a lower genus remeshing of $S(x,y,z)$ with approximately equilateral triangle faces. If the input $S(x,y,z)$ is a smooth shape with only small holes or handles such as the cMCF intermediary $S_{\text{ref}}(x,y,z)$ the proposed voxelization and remesh yields a genus-0 mesh.

*Step 2: $S_{\text{ref}}(x,y,z)$ to $S_Q^2(x,y,z)$*

*Quasi-conformal spherical parametrization of genus-0 closed surfaces.* For genus-0 closed surfaces the uniformization theorem[52] guarantees the existence of a conformal map onto the unit sphere, $\mathbb{S}^2$. For a closed orientable surface such as $S_{\text{ref}}(x,y,z)$ we can compute the genus, $g$ from the Euler characteristic, $\chi = 2 - 2g = \#V - \#E + \#F$. If $g = 0$ we applied the method of Choi et al.[39,51] which uses the theory of quasi-conformal composition to ensure a bijective spherical parametrization with bounded conformal error (i.e. quasi-conformal). In practice, we found conformal errors = 0 (Extended Fig. 3).

*Step 3: $S_Q^2(x,y,z)$ to $S_\Omega^2(x,y,z)$*

*Equiareal spherical parameterization by mesh relaxation.* We iteratively advect the vertex coordinates of $S_Q^2(x,y,z)$, whilst preserving face connectivity and the spherical shape to minimise the per face area distortion factor, $\lambda$. The magnitude and direction to advect each vertex, the vector field $\vec{V}$ was found as the solution to the linear heat equation[27], $\frac{d\lambda}{dt} = -\Delta\lambda$, where $\Delta$ denotes the Laplacian. This is because an infinitesimal change in $\lambda$ in the direction of $\vec{V}$ is the Lie-derivative on 2-forms, $\mathcal{L}_{\vec{V}}\lambda = -\nabla \cdot (\lambda\vec{V})$ such that we can set $\Delta\lambda = -\nabla \cdot (\lambda\vec{V})$. As the Laplacian is the divergence of the gradient, $\Delta\lambda = \nabla \cdot (\nabla\lambda)$ we have $\vec{V} = -\frac{\nabla\lambda}{\lambda} = -\nabla\log\lambda$. To ensure vertices are displaced geodesically on the surface of a sphere according to $\vec{V}$, instead of unwrapping the sphere to the 2D $(u,v)$ plane as in Lee et al[27], we developed a direct 3D advection scheme that displaces vertices in small constant step sizes $\epsilon$ using active contour cMCF (see below), and reprojecting to the sphere. $\lambda$ and $\vec{V}$ are recalculated for the new vertex positions and advection is repeated until an equiareal parameterization was achieved or the maximum number of iterations was reached. Details of our advection scheme is given algorithmically.

**Input:** Conformal spherical parameterization mesh, $S_Q^2(x,y,z)$ with vertices, $v_{sphere}$ and faces, $f_{sphere}$, matching genus-0 mesh, $S_{ref}(x,y,z)$ with vertices, $v_{ref}$ and faces, $f_{ref}$ with identical number of vertices, $|v_{sphere}| = |v_{ref}|$, and faces, $f_{sphere} = f_{ref}$; vertex step size, $\epsilon$; total number of iterations, $T$; mesh stiffness factor, $\delta$ (also known as the time step, $\delta$ in cMCF)

**For** iterations $t = 1,2,\ldots,T \ldots$ do :

$a_{ref}^{f_i} \leftarrow a_{ref}^{f_i}/\Sigma_i a_{ref}^{f_i}$      (normalised area per face $f_i$ of $S_{ref}$)

$a_{sphere}^{f_i}(t) \leftarrow a_{sphere}^{f_i}(t)/\Sigma_{f_i} a_{sphere}^{f_i}(t)$      (normalised area per face $f_i$ of $S_Q^2$)

$\lambda^{f_i}(t) \leftarrow a_{ref}^{f_i}/a_{sphere}^{f_i}(t)$      (area distortion factor per face $f_i$ of $S_Q^2$)

$\vec{V}^{f_i}(t) = -\nabla\log\lambda^{f_i}(t)$      (compute $\nabla$ using the mesh grad operator[106], per face $f_i$)

$\vec{V}^{f_i}(t) \leftarrow \left(\frac{\bar{l}}{median(\|\vec{V}^{f_i}(t)\|)}\right)\vec{V}^{f_i}(t)$      (normalize the displacement vector with respect to the average triangle edge length, $\bar{l}$)

$\lambda^{v_i}(t) \leftarrow average(\lambda^{f_i}(t))$,
$\vec{V}^{v_i}(t) \leftarrow average(\vec{V}^{f_i}(t))$      (average the per face $f_i$ vector field and area distortion factor onto vertices, $v_i$)

$$\vec{V}^{v_i}_{tangent}(t) \leftarrow \vec{V}^{f_i}(t) - \left(\vec{V}^{f_i}(t) \cdot \vec{N}^{v_i}(t)\right)\vec{N}^{v_i}(t)$$
(compute the surface tangential component of $\vec{V}^{v_i}(t)$ using per vertex normal, $\vec{N}^{v_i}(t)$)

$$v_{sphere}(t+1) = (\mathbf{M}(t) - \delta_t \mathbf{L}(t))^{-1}\mathbf{M}(t)\left(v_{sphere}(t) + \epsilon \vec{V}_{tangent}(t)\right)$$
(perform active contour cMCF to advect $v_{sphere}$ in direction of $\vec{V}^{v_i}_{tangent}(t)$ and compute the new $v_{sphere}$)

$$v_{sphere}(t+1) \leftarrow v_{sphere}(t+1)/\|v_{sphere}(t+1)\|$$
(Ensure $v_{sphere}$ lies on a sphere by centroid distance normalization)

This mesh relaxation bijectively diffuses the area distortion scalar factor on the sphere surface in a stable manner until a triangle face collapses, that is when an interior angle = 0. The extent of area relaxation is determined by the input mesh quality. We find that if $S_{\text{ref}}(x,y,z)$ is regular, with near-equilateral faces and has a minimal number of vertices (> 40k for a pixel resolution of 0.104μm), $S_Q^2(x,y,z)$ could be stably relaxed to an equiareal spherical parameterization, $S_\Omega^2(x,y,z)$, without triangle collapse.

*Step 4:* $S_\Omega^2(x,y,z)$ to $S(u,v)$

*Automatic determination of unwrapping axis using weighted principal components analysis (PCA).* Mapping surface features of interest with minimal geometrical distortion into $S(u,v)$ is equivalent to finding an optimal north-south unwrapping axis for $S_\Omega^2(x,y,z)$. This optimization is solved by weighted PCA. Let $v_i = (x_i, y_i, z_i)$ denote the coordinate of vertex $i$ on $S_\Omega^2(x,y,z)$ and $w_i$ the vertex weight, a score of the importance of mapping this vertex with minimum geometrical distortion. The 3x3 weighted covariance matrix, $\mathbf{A} = (\mathbf{w}^T \mathbf{v})(\mathbf{w}^T \mathbf{v})^T$ over all vertices captures the spread of the weight over the sphere. Eigendecomposition applied to the symmetric matrix $\mathbf{A}$ finds the principal orthogonal directions of variance given by eigenvalues $\boldsymbol{\lambda} = [\lambda_1, \lambda_2, \lambda_3]$, $\lambda_1 \geq \lambda_2 \geq \lambda_3$ and eigenvectors $\mathbf{e} = [\mathbf{e_1}, \mathbf{e_2}, \mathbf{e_3}]$. The eigenvalue captures the concentration of the weight $w$ in the direction of the corresponding eigenvector. The optimal north-south unwrapping axis is the smallest eigenvector, $\mathbf{e_3}$. To unwrap with respect to $\mathbf{e_3}$ we rotate the vertex coordinates $v_i$ so that $\mathbf{e_3}$ is the new z-axis. As the eigenvector matrix $\mathbf{e}$ is orthonormal and thus a 3D rotation matrix, $\mathbf{e}$ is the rotation matrix $\mathbf{R}$ that maps the x-axis, $\begin{pmatrix}1\\0\\0\end{pmatrix} \mapsto \mathbf{e_1}$, y-axis, $\begin{pmatrix}0\\1\\0\end{pmatrix} \mapsto \mathbf{e_2}$, z-axis, $\begin{pmatrix}0\\0\\1\end{pmatrix} \mapsto \mathbf{e_3}$. For a 'pure' or proper rotation matrix without reflection the determinant of $\mathbf{R}$ must be +1, $\det(\mathbf{R}) = +1$. We derive a proper rotation, $\mathbf{R}' = [\mathbf{e_1}', \mathbf{e_2}', \mathbf{e_3}']$ from $\mathbf{R}$ by flipping the sign of $\mathbf{e_1}, \mathbf{e_2}$ to have positive x- and y- components respectively; $\mathbf{e_1'} \leftarrow sgn(v_{1x})\mathbf{e_1}$, $\mathbf{e_2'} \leftarrow sgn(v_{2y})\mathbf{e_2}$ (sgn is the sign function) and constructing $\mathbf{e_3'} = \mathbf{e_1'} \times \mathbf{e_2'}$ as the cross product of $\mathbf{e_1'}$ and $\mathbf{e_2'}$. The matrix inverse of $\mathbf{R}'$ (also the matrix transpose, $\mathbf{R'^T}$) is the desired rotation of $S_\Omega^2(x,y,z)$ such that $\mathbf{e_3'}$ is the new z-axis. $\mathbf{R'^T}$ maps the eigenvectors, $\mathbf{e_1'} \mapsto \begin{pmatrix}1\\0\\0\end{pmatrix}$, $\mathbf{e_2'} \mapsto \begin{pmatrix}0\\1\\0\end{pmatrix}$, $\mathbf{e_3'} \mapsto \begin{pmatrix}0\\0\\1\end{pmatrix}$.

*UV-mapping the unit sphere.* We construct an equidistant UV unwrap of the unit sphere where $u$, the column coordinate equidistantly samples the circumference of the sphere, a total length $2\pi$ and $v$, the row coordinate equidistantly samples the arc from north to south pole, a total length $\pi$. This specifies a $N \times 2N$ pixel UV image with $N$ as a user-defined size. By default $N = 256$ pixels. The UV mapping is constructed by pullback. Let $u = \theta$ be the azimuthal and $v = \varphi$ be the inclination angles of the sphere and setup the $N \times 2N$ grid of $v$ vs $u$ over the parameter space $[-\pi, 0] \times [-\pi, \pi]$. Convert the spherical coordinates, $(1, \theta, \varphi)$ to cartesian coordinates, $(x,y,z) = (\sin\theta\cos\varphi, \sin\theta\sin\varphi, \cos\theta)$. Each $(x,y,z)$ coordinate is matched by nearest distance to a triangle face, $ABC$ of $S_\Omega^2(x,y,z)$ to compute barycentric coordinates giving $(x,y,z)$ as a convex combination of the vertices $A, B, C$; $\mu_A A + \mu_B B + \mu_C C$, $\mu_A, \mu_B, \mu_C \geq 0$ and $\mu_A + \mu_B + \mu_C = 1$. By bijectivity of $S_{\text{ref}}(x,y,z)$ and $S_\Omega^2(x,y,z)$, we set $A, B, C$ to the respective $S_{\text{ref}}(x,y,z)$ vertex coordinates to produce the respective uv- coordinate mapping, $(u,v) \leftrightarrow \mu_A A + \mu_B B + \mu_C C$. Note, setting $A, B, C$ to be the vertex coordinates of any mesh bijective to $S_\Omega^2(x,y,z)$ e.g. $S(x,y,z)$ in the direct unwrapping case produces similarly the corresponding uv- coordinate mapping for that mesh. The weights $\mu_A, \mu_B, \mu_C$ is also used to map any other vertex associated quantities, $I_i(S_\Omega^2(x,y,z))$ such as curvature to $I_i(S(u,v))$, with vector-valued vertices $A, B, C$ replaced now by scalar values. The construction of the UV map as described above replicates the first and last column of the resulting UV image. For applications such as texture mapping and active contour cMCF (see analysis of ruffles in Datasets section) where the UV image grid needs to be converted to a triangular mesh and the image boundaries 'stitched' together we instead use a $N \times 2N + 1$ pixel UV image.

*Step 5:* $V(x, y, z)$ to $V(d, u, v)$

*Topographic coordinate space $(d, u, v)$ construction.* UV-unwrapping establishes bijection between a 2D uv plane and a 3D Cartesian surface, $(u, v) \leftrightarrow S_{\text{ref}}(x, y, z)$. We construct a topographic $(d, u, v)$ coordinate space, $V(d, u, v)$ corresponding to a volume space normal to $S_{\text{ref}}(x, y, z)$ by propagating the $(u, v)$ parameterized $S_{\text{ref}}(x, y, z)$ in Cartesian 3D at equidistant steps of $\alpha$ voxels, referred to as $\alpha$-steps, along the steepest gradient of the signed distance function, $\nabla\Phi(x, y, z)$ for a total of $D$ steps, $d \in -D_{in}, \ldots, D_{out}$. $D = D_{out} + D_{in}$ is the total number of $\alpha$-steps outwards and inwards relative to $S_{\text{ref}}(x, y, z)$ (which is $d = 0$) respectively. $D_{out}$ is automatically determined to ensure $V(d, u, v)$ fully encapsulates $S(x, y, z)$. $D_{in}$ is user-defined for computational efficiency or automatically determined as a fraction of the maximum internal distance transform. The signed distance function $\Phi(x, y, z)$ of $S_{\text{ref}}(x, y, z)$ is computed from the binary volume after voxelization. We voxelize the $(u, v)$ parameterized $S_{\text{ref}}(x, y, z)$ directly using the same procedure as for meshes but employ image upscaling instead of mesh subdivision to ensure that the distance of 1 pixel in the $(u, v)$ space is < 1 voxel in Cartesian $(x, y, z)$ space. Using active contour cMCF (see below) to propagate the $(u, v)$ parameterized $S_{\text{ref}}(x, y, z)$ for large $D$ is slow; a 256x512 UV unwrap is 131,072 vertices. Moreover for a large $D_{out}$ as the intra-spacing of 3D $(x, y, z)$ positions increases, numerical instabilities arise that require implicit Laplacian smoothing[107] to suppress, which is also slow. Instead we use explicit Euler integration for propagation; $S_{\text{ref}}(x, y, z)_{d+\alpha} = S_{\text{ref}}(x, y, z)_d + \alpha \frac{\nabla\Phi_{S_{\text{ref}}(x,y,z)_d}}{|\nabla\Phi_{S_{\text{ref}}(x,y,z)_d}|}$ at $\alpha$ voxels from $S_{\text{ref}}(x, y, z)_d$ and $\frac{\nabla\Phi_{S_{\text{ref}}(x,y,z)_d}}{|\nabla\Phi_{S_{\text{ref}}(x,y,z)_d}|}$ is the unit gradient of $\Phi(x, y, z)$. Computationally efficient image-based filtering is then applied to smooth $S_{\text{ref}}(x, y, z)_{d+\alpha}$ per iteration to maintain bijectivity and suppress instabilities. Tilinear interpolation of the respective Cartesian volumetric signal intensities, $I_i(V(x, y, z))$ at the $(x, y, z)$ coordinates indexed by $V(d, u, v)$ generates the topographic 3D equivalents, $I_i(V(d, u, v))$.

*Step 6:* $I_i(V(d, u, v))$ to $S(d, u, v)$
*Topographic mesh $S(d, u, v)$ construction.* $S(x, y, z)$ was voxelized to a binary volume, $I_i(V(x, y, z))$ as above and transformed to $I_i(V(d, u, v))$. Marching cubes were applied at isovalue = 0.5 to create an initial $S(d, u, v)$ which was remeshed with ACVD to construct the final low-genus $S(d, u, v)$ with near-equilateral triangle faces. The Cartesian 3D mesh, $S_{\text{topo}}(x, y, z)$ of $S(d, u, v)$ was constructed by interpolation of the $(x, y, z)$ coordinates indexed by the corresponding $(d, u, v)$ coordinates. To transform surface signals, $F_i(S(x, y, z))$ to $F_i(S(d, u, v))$, nearest neighbors was used to match $S(x, y, z)$ and $S_{\text{topo}}(x, y, z)$.

Mesh displacement by active contour cMCF
Active contours, or 'snakes'[108], define the boundary of an image region by minimizing its contour energy, $E$. The contour energy is the sum of an internal, $E_{int}$ and an external energy, $E_{image}$, $E(v, I) = E_{image}(v, I) + E_{int}(v)$. The internal energy is set by $E_{int} = \int \alpha|v'|^2 + \beta|v''|^2 \, ds$, where the number of ' denotes the order of the spatial derivative. Here, the first term is the tension and $\alpha$ the elasticity of the contour. The second term is the stiffness and $\beta$ the rigidity of the contour. The external energy is set by $E_{image} = -\int p \, ds$, where $p$ is an attractor image for the contour. Minimizing $E$ is equivalent to solving the Euler-Lagrange equation, $\alpha v'' - \beta v'''' = -\nabla p$ or in matrix form, $\mathbf{A}v + \nabla p = 0$, where $\mathbf{A}$ prescribes the constant coefficients for computing the second and fourth order derivatives by finite differences. Given a vertex position $v(t)$, the next position, $v(t + 1)$ is computed that minimises the residual error $\mathbf{A}v + \nabla p$ using gradient descent and backwards Euler is $v(t + 1) = v(t) - (\mathbf{A}_t v(t + 1) + \nabla p)$ and the linear system is $(\mathbf{I} + \mathbf{A}_t)v(t + 1) = v(t) + \nabla p$, where $\mathbf{I}$ is the identity matrix. If $\beta = 0$, $\mathbf{A}_t$ only comprises the second order coefficients associated with $v''$, and we have $(\mathbf{I} + \alpha_t \mathbf{M}_t^{-1} \mathbf{L}_t)v(t + 1) = v(t) + \nabla p$ or equivalently, $(\mathbf{M}_t - \alpha_t \mathbf{L}_t)v(t + 1) = \mathbf{M}_t(v(t) + \nabla p)$ which is identical to the cMCF equation with $\mathbf{M}_t$ the mass matrix, $\mathbf{L}_t$, the Laplacian matrix, $\alpha_t = \delta_t$ and $\mathbf{L}_t = \mathbf{L}_0$ in response to an external force, $\nabla p$. This general equation can be solved by direct matrix inversion to move surface meshes diffeomorphically. We refer to this as *active contour cMCF* in this paper. To evolve mesh vertices, $v(t)$ normal to the surface in equal $\alpha$-steps, we set $p = \Phi$, the signed distance function, and solve $(\mathbf{M}_t - \alpha_t \mathbf{L}_t)v(t + 1) = \mathbf{M}_t(v(t) + \alpha \nabla \Phi)$ iteratively, with $\nabla \Phi$ evaluated at $v(t)$ for each iteration. A positive $\alpha$ moves a cell surface mesh normally outwards from the cell and a negative $\alpha$ moves the mesh normally into the cell.

Quantification of geometric deformation errors for meshes
Surface mappings do not conserve local geometrical measures like angles, edge lengths and face area. Quantification of the distortion in these measures enables a task-specific optimization of the mapping and correction of statistical measurements made on the mapped surface. There are two primary geometric distortions to quantify; conformal and area distortion error (Extended Fig. 1). An isometric deformation is one with no error; both conformal and area distortion errors are 0.

*Conformal error*. The conformal or quasi-conformal error, $Q_i$ measures the extent the shape of a mesh element $i$, e.g. a triangle face, is stretched. It is 0 if the relative distances between vertices and the angles between edges are preserved after the mapping. We compute $Q$ of mapping triangle $\Delta ABC$ to $\Delta DEF$ in 3D by first isometrically projecting all triangles into 2D. Let $A = (x_1, y_1, z_1)$, $B = (x_2, y_2, z_2)$, $C = (x_3, y_3, z_3)$ with edge vectors, $\overrightarrow{AB} = B - A$, $\overrightarrow{AC} = C - A$, then an identical 2D triangle $\Delta A'B'C'$ is given by $A' = (0,0)$, $B' = (|\overrightarrow{AB}|, 0)$, $C = (\overrightarrow{AB} \cdot |\overrightarrow{AB} \times \overrightarrow{AC}|)$, $A', B', C' \in \mathbb{R}^2$. Let $\mathbf{X} = \begin{bmatrix} A' & B' & C' \\ 1 & 1 & 1 \end{bmatrix}$, be the 3x3 homogeneous vertex coordinates of $\Delta A'B'C'$ and $\mathbf{Y} = \begin{bmatrix} D' & E' & F' \\ 1 & 1 & 1 \end{bmatrix}$, the 3x3 homogeneous vertex coordinates of $\Delta D'E'F'$ then we solve for the 3x3 matrix, $\mathbf{A}$ that maps $\mathbf{X}$ to $\mathbf{Y} = \mathbf{AX}$. $\mathbf{A}$ is affine and of the form $\begin{bmatrix} \mathbf{J} & | & \mathbf{T} \\ \mathbf{0} & | & 1 \end{bmatrix}$ where $\mathbf{J}$ is a 2x2 transformation matrix and $\mathbf{T}$ a translation matrix. Eigenvector decomposition of $\mathbf{J}^T\mathbf{J}$ gives 2 eigenvalues $\lambda_1, \lambda_2$, $\lambda_1 < \lambda_2$ and the singular values of $\mathbf{J}$, $\sigma_1 = \sqrt{\lambda_1}, \sigma_2 = \sqrt{\lambda_2}$. The ratio $\frac{\sigma_2}{\sigma_1}$ is the conformal error[62]. The global conformal error, $Q$ of deforming a surface mesh $S_1$ to a mesh $S_2$ is the area weighted average of individual conformal errors $Q_{f_i}$ of each triangle face $f_i$ in $S_1$; $Q = \frac{\Sigma_{f_i \in S_1} a_{f_i} Q_{f_i}}{\Sigma_{f_i \in S_1} a_{f_i}}$ where $a_{f_i}$ is the area of face $f_i$ of $S_1$.

*Area distortion error*. The area distortion error, $\lambda$ measures the extent the surface area fraction of a mesh face is preserved during a surface mapping. One measure of $\lambda$ is $\sigma_1 \sigma_2$, the product of the singular values of $\mathbf{J}$ and the area of the distortion ellipse[63]. Here we use the surface area fraction ratio, $\lambda_{\Delta ABC} = \frac{\frac{area(\Delta ABC)}{surface\ area\ of\ S_1}}{\frac{area(\Delta DEF)}{surface\ area\ of\ S_2}}$ as a direct measurement of the area distortion in mapping $\Delta ABC$ to $\Delta DEF$ in 3D. The global area distortion error, $\lambda = \frac{1}{|f_i|} \Sigma_{f_i \in M_1} \lambda_{f_i}$ for mapping a mesh $S_1$ to a mesh $S_2$ is the mean over all individual area distortion $\lambda_{f_i}$ of each triangle face $f_i$ in $S_1$, with $|f_i|$ the number of faces in $S_1$ and $\lambda_{f_i} = \frac{a_{f_i}^{S_1}/\Sigma_{f_i \in S_1} a_{f_i}^{S_1}}{a_{f_i}^{S_2}/\Sigma_{f_i \in S_2} a_{f_i}^{S_2}}$ is the area distortion of face $f_i$. The normalization of face area by total surface area is crucial to enable the computation of $\lambda$ independent of scale.

Quantification of geometric deformation error for UV images
UV mapping defines a bijective relation between the 2D $(u, v)$ rectilinear grid and a 3D surface, $S(u, v) \leftrightarrow S = S(x(u,v), y(u,v), z(u,v))$. Differentials can be used to compute geometric quantities when the $(u, v)$ spacing is comparable to the $(x, y, z)$ spacing. The differential area of a $(u, v)$ pixel is $dA = \left|\frac{\partial S}{\partial u} \times \frac{\partial S}{\partial v}\right| du\, dv$, where $\frac{\partial S}{\partial u} = \left(\frac{\partial S_x}{\partial u}, \frac{\partial S_y}{\partial u}, \frac{\partial S_z}{\partial u}\right)$ and $\frac{\partial S}{\partial v} = \left(\frac{\partial S_x}{\partial v}, \frac{\partial S_y}{\partial v}, \frac{\partial S_z}{\partial v}\right)$ are the image gradients of the $x, y, z$ surface coordinates in $u, v$ directions. The topographic space construction establishes bijection of the 3D $(d, u, v)$ volumetric grid to a 3D volume, $V$, $(u, v, d) \leftrightarrow V = V(x(u,v,d), y(u,v,d), z(u,v,d))$. The differential volume of a $(d, u, v)$ voxel is $dV = \left|\left(\frac{\partial V}{\partial u} \times \frac{\partial V}{\partial v}\right) \cdot \frac{\partial V}{\partial d}\right| du\, dv\, dd$. The matrix $\left[\frac{\partial S}{\partial u}, \frac{\partial S}{\partial v}\right]$ is the 2x3 Jacobian matrix, $\mathbf{J}$ and the conformal error per pixel is $\frac{\sigma_2}{\sigma_1}$, $\sigma_1 = \sqrt{\lambda_1}, \sigma_2 = \sqrt{\lambda_2}$ where $\lambda_1, \lambda_2, \lambda_1 < \lambda_2$ are the two eigenvalues of $\mathbf{J}^T\mathbf{J}$. The global conformal error $Q$ of $uv$-mapping the surface mesh $S$ is the differential area weighted average of individual conformal errors $Q_{uv}$ of each $uv$ pixel; $Q = \frac{\Sigma_{uv} dA_{uv} Q_{uv}}{\Sigma_{uv} dA_{uv}}$ where $dA_{uv}$ is the area element of the $uv$ pixel. The area distortion error per $uv$ pixel is the ratio between the surface area fraction of a $uv$ pixel and the corresponding surface element on $S$, $\lambda_{uv} = \frac{\frac{dudv}{\Sigma_{uv} dudv}}{\frac{dA_{uv}}{\Sigma_{uv} A_{uv}}}$. Note $dudv = 1$ and $\Sigma_{uv} dudv =$ total number of $uv$ pixels. The global area distortion error, $\lambda = \frac{1}{|uv|} \Sigma_{uv} \lambda_{uv}$ for $uv$ mapping a surface $S$ is the mean over all individual area distortion $\lambda_{uv}$ of each $uv$ pixel.

Stopping criteria for area distortion relaxation of $S_Q^2(x, y, z)$
We used three additional stopping criteria to demonstrate intermediate area distortion relaxation between fully conformal, $S_Q^2(x, y, z)$ and fully equiareal, $S_\Omega^2(x, y, z)$ parameterization. We use the same nomenclature as for the above discussed geometric deformation error for meshes.

*Most isometric parametrization (MIPS) error*. The MIPS[62] error is defined $\frac{\sigma_2}{\sigma_1} + \frac{\sigma_1}{\sigma_2}$ and is minimal when $\sigma_1 = \sigma_2$. This error is trivially minimal for a conformal spherical parametrization $\left(\frac{\sigma_1}{\sigma_2} = 1\right)$ (Extended Fig. 3b).

*Area-preserving MIPS.* The area-preserving MIPS[63] is defined $\left(\frac{\sigma_1}{\sigma_2} + \frac{\sigma_2}{\sigma_1}\right)\left(\sigma_1\sigma_2 + \frac{1}{\sigma_1\sigma_2}\right)^\theta$. We use this metric with $\theta = 1$, which measures the uniformity of stretch distortion over a surface. This error yields near-equiareal spherical parametrization (Extended Fig. 3b).

*Isometric error.* We observed tradeoff of conformal error, $Q$ and log area distortion, $\log \lambda$ on a similar magnitude scale such that their summation has a unique global minima (Extended Fig. 1e). We thus define an isometric error metric, $(1-\theta)\frac{\sigma_1}{\sigma_2} + (\theta)\log\lambda$ with a constant $\theta \in [0,1]$ to weight the relative importance of jointly minimizing conformal and area distortion error. We use $\theta = 0.5$ in Extended Fig. 3b.

Assessment of geometrical difference between two meshes

Four metrics were used to assess the difference between two meshes $S_1$ and $S_2$ possessing different number of vertices and faces; chamfer distance (CD), Wasserstein-1 distance ($W_1$), the difference in surface area ($\Delta A$) and the difference in volume ($\Delta V$). CD is the mean Euclidean distance between all vertices of $S_1$ when matched to the nearest vertex of $S_2$ and vice versa, $CD = \frac{1}{|S_1|}\sum_{x \in S_1}\min_{y \in S_2}\|x-y\|_2 + \frac{1}{|S_2|}\sum_{y \in S_2}\min_{x \in S_1}\|x-y\|_2$. The Wasserstein-1 distance ($W_1$) or Earth-mover's distance (EMD) is the minimum total area weighted distance of 1-to-1 matching vertices on $S_1$ and $S_2$. $W_1$ accounts for the area of triangle faces and is minimal if the vertices of $S_1$ is a uniform sampling of $S_2$ or vice versa. Exact computation of $W_1$ is impractical, even for small meshes. We compute $W_1$ using the sliced-Wasserstein, $SW_1$ approximation, which uses random spherical projections to sum multiple 1D EMD distances[64]. Specifically we use the *ot.sliced.max_sliced_wasserstein_distance* function from the Python *POT* library with 50 projections and average the result from 10 evaluations to report an estimate. The difference in total surface area is $\Delta A = A_{S_1} - A_{S_2}$ and is $\Delta A(\%) = \frac{A_{S_1} - A_{S_2}}{A_{S_2}}$ when given as a percentage. Total surface area was computed as the sum of individual triangle areas. The difference in total volume is $\Delta V = V_{S_1} - V_{S_2}$ and is $\Delta V(\%) = \frac{V_{S_1} - V_{S_2}}{V_{S_2}}$ when given as a percentage. The volume of a mesh was computed as the number of voxels in its binary voxelization computed as described above. We used the minimal possible dilation ball kernel size to ensure a correct volume computation - visual checking of binary voxelization and value at least 3x surface area. We use $S_1 = S_{\text{topo}}(x,y,z)$ and $S_2 = S(x,y,z)$ to compute the metrics of Extended Fig. 2,3.

Reference surface inference for measurement of protrusion height

An optimal reference surface for protrusion segmentation must be a $(u,v)$ parameterized surface i.e. $S_{\text{ref}}(d_{\text{ref}} = f(u,v), u, v)$ where $f(\cdot)$ is injective such that every surface point is defined by a unique $(d, u, v)$-tuple. We prove this by contradiction. Suppose a surface, $S(d, u, v)$ has points with the same $(u, v)$ but different $d$ coordinates. The points with higher $d$ must therefore be part of a surface protrusion and thus $S(d, u, v)$ cannot be a $S_{\text{ref}}(d_{\text{ref}}, u, v)$. A suitable $S_{\text{ref}}(d_{\text{ref}}, u, v)$ can thus be found as the $(u, v)$ parametrized 2D 'baseline' surface, $d_{\text{ref}} = f_{\text{smooth}}(u, v)$ to a 2D adaptation of the asymmetric least squares problem (ALS)[109]; $d_{\text{ref}} = \arg\min_z \left\{\sum(w_{uv}(d_{uv}^{S'(d,u,v)} - d_{uv})^2 + \lambda\sum_{uv}(\Delta d_{uv})^2\right\}$ with asymmetric weights, $w_{uv}$: $w_{uv} = p$ if $d_{uv}^{S'(d,u,v)} > d_{uv}$ and $w_{uv} = 1 - p$ otherwise. The regularization parameter, $\lambda$ controls the contribution of the Laplacian $\Delta d_{uv} = \nabla^2 d_{uv}$. The solution is a surface intermediate between a $(u, v)$-parameterization approximation, $d_{uv}^{S'(d,u,v)} = f(u,v)$ of the topographic surface $S(d \approx f(u,v), u, v)$ and the flat 2D-plane ($d = 0$) (Extended Fig. 3a). The input 1024x512 pixels approximation ($d_{uv}^{S'(d,u,v)}$) was computed as an image by extending a vertical ray upwards at each $(u, v)$ pixel and setting the image pixel value as the longest contiguous stretch of the topographic binary. We downsample $d_{uv}^{S'(d,u,v)}$ 8x to 128x64 pixels for computational efficiency and additional smoothness regularization and solve for $d_{\text{ref}}$ by running 10 iterations of ALS[109] using $p = 0.25, \lambda = 1$. The solution, $d_{\text{ref}}$ is resized back to 1024x512 pixels. The height, $h$ of $S(d, u, v)$ relative to the inferred reference surface is the difference, $h = d - d_{ref}$ between a vertex's $d$ coordinate and $d_{\text{ref}}$ of the matching point on $S_{\text{ref}}(d_{\text{ref}} = f_{\text{smooth}}(u,v), u, v)$ as found by interpolation.

Topography guided binary segmentation of protrusions

For cMCF binary segmentation of $S(d,u,v)$, the reference surface used is the 2d $(u,v)$ plane, $S_{\text{ref}}(d_{\text{ref}},u,v) = S(d=0,u,v)$ and the height is $h = d$. For more optimal segmentation, the reference surface, $S_{\text{ref}}(d_{\text{ref}},u,v)$ was inferred as above and the height is $h = d - d_{\text{ref}}$, relative to the matching point on $S_{\text{ref}}(d_{\text{ref}},u,v)$ with identical $(u,v)$ coordinate. For both, the mean height, $\bar{h}$ is the threshold to give the initial binary segmentation, $F_i(S(d,u,v)) = h \geq \bar{h}$. We postprocess by applying graph connected component analysis to remove small segmented regions with surface area < 200 voxels$^2$; diffusing the result using two-class labelspreading[110] with an affinity matrix, $A$, for 20 iterations, clamping ratio 0.99, and binarizing the label probability with threshold of 0.25 at the start of each iteration. Lastly, any remaining small regions with surface area < 500 voxels$^2$ was removed. The affinity matrix[69], $A$ is a weighted sum ($\gamma = 0.9$) of an affinity matrix based on geodesic distance, $A_{dist}$ and one based on surface convexity, $A_{convex}$; $A = \gamma A_{dist} + (1-\gamma) A_{convex}$ of $S(d,u,v)$. $A_{dist} = \begin{cases} e^{-D_{dist}^2/(2\mu(D_{dist})^2)} & i \neq j \\ 1 & i = j \end{cases}$ where $D_{dist}$ is the pairwise Euclidean distance matrix between two vertices $i$ and $j$. $A_{convex} = \begin{cases} e^{-D_{convex}^2/(2\mu(D_{convex})^2)} & i \neq j \\ 1 & i = j \end{cases}$ where $D_{convex}$ is the pairwise Cosine distance, $\frac{(1-\cos(\theta_{ij}))}{2}$ matrix of the dihedral angle, $\theta_{ij}$ between the normal vectors at two vertices $i$ and $j$. $\mu(D)$ denotes the mean value of the entries of matrix $D$.

### Topography guided instance segmentation of protrusions

Individual protrusions are segmented by identifying high curvature protrusive features and applying connected components analysis. We compute the topographic mean curvature $H(S(d,u,v)) = -\frac{1}{2}\nabla \cdot \hat{n}$ with the normal, $\hat{n}$ given by the unit gradient of the signed distance transform of the binary topographic volume of the cell, $I_{binary}(V(d,u,v))$. We compute a binary subvolume restricted to the surface, $I_{surf}(V(d,u,v))$, the intersection of the morphological dilation of $I_{binary}(V(d,u,v))$ with ball kernel size 2, and the morphological erosion of $I_{binary}(V(d,u,v))$ with ball kernel size 2. To identify high curvature surface regions, $H_{high}(d,u,v)$ for lamellipodia, we concatenate $H(S(d,u,v))$ Gaussian smoothed with $\sigma = 1,3,5$ as a 3-dimensional feature for all voxels in $I_{surf}(V(d,u,v))$ and apply Gaussian mixture model (GMM) clustering (# classes = 3), keeping the class with the highest mean $H$. To identify $H_{high}(V(d,u,v))$ for blebs and filopodia which are circular and smaller, we use $H(S(d,u,v))$ Gaussian smoothed with $\sigma = 1$ as a 1-dimensional feature for all voxels in $I_{surf}(V(d,u,v))$ and apply kmeans clustering (# classes = 3), keeping the class with the highest mean $H$. For efficiency, both GMM and kmeans clusterers are fitted on a random sampling of 10,000 surface voxels. Small regions with < 500 connected voxels are removed. Connected component analysis labels each disconnected region in $H_{high}(V(d,u,v))$ as individual protrusions, $I_{protrusions}(V(d,u,v))$. We expand labels by 3 voxels and transfer the segmentation to the surface mesh, $S(d,u,v)$ by interpolation at the vertex coordinates, $F_{protrusions}(S(d,u,v))$ for further surface-based processing. We first apply the binary protrusion segmentation above, $B_{protrusion}(S(d,u,v))$ to $F_{protrusions}(S(d,u,v))$, taking the intersection and keeping segmentations with size > 100 voxel$^2$ Cartesian 3D surface area. We diffuse segmentation labels with labelspreading, clamping ratio 0.99 for 10 iterations, with affinity matrix $A$, $\gamma = 0.9$ as above. We do not rebinarize the label probability at the start of each iteration. Finally, we apply $B_{protrusion}(S(d,u,v))$ to the diffused segmentations to get the final instance segmentation labels, $F_{protrusions}(S(d,u,v))$.

### Direct 2D unwrapping of protrusion submeshes

Segmented individual protrusions are open 3D surfaces with disk topology and can be directly unwrapped into 2D if they possess no holes or handles and have one boundary. The genus, $g$ of an open orientable surface with $b$ boundaries is computed from the Euler characteristic, $\chi = 2 - 2g - b = \#V - \#E + \#F$. Similar to the spherical parameterization of closed 3D surfaces, the open 3D surface is first mapped conformally to the unit disk then relaxed to get an equiareal disk parameterization.

*Quasi-conformal disk parametrization of genus-0 open surfaces.* We obtain a quasi-conformal map of an open 3D surface to the unit disk by harmonic parametrization[111]. The boundary vertices are first mapped to the boundary of the unit circle, whilst preserving edge length fractions. Interior vertices are then mapped to the disk interior by solving Laplace's equation, $\nabla^2 \phi = 0$.

*Equiareal disk parameterization by mesh relaxation.* We relax the conformal disk parametrization whilst preserving the boundary topology using the area-preserving flow method[112]. We solve Poisson's equation to compute the smooth vector field for diffusing the area distortion and explicit Euler integration to advect vertex points iteratively with Delaunay triangle flips. The extent of area relaxation achieved is determined by the mesh quality and number of vertices with respect to the extremity of local area distortion. In general, relaxation was less stable compared to our relaxation for spherical surfaces above. For thin and long protrusions, prior downsampling and uniform remeshing of the protrusion submesh was necessary to enable full area distortion relaxation.

To convert a unit disk parameterization to an $N \times N$ pixel image, we 'square' the disk using the elliptical grid mapping formula[113], multiply the resulting vertex coordinates by $N/2$ and interpolate the coordinates and associated vertex quantities onto a $N \times N$ pixel integer grid. This gives similar results to but is significantly faster than solving the Beltrami equation[94].

Refinement of undersegmented blebs

Given $S(x, y, z)$ and the vertex ids corresponding to protrusion $i$, $(v_i)$ we first impute any small holes in the segmentation; inner vertices not assigned to protrusion $i$ but should be in order to ensure the protrusion submesh, $S_{\text{protrusion}}(x, y, z)$ is a genus-0 open surface. We do this by applying graph connected component analysis on the submesh formed by all vertex ids not part of protrusion $i$, $\{v\}\backslash\{v\}^i$. Any component with number of vertices <10% the total surface area of $S(x, y, z)$ is assigned to protrusion $i$ to form $\{v\}^i_{\text{impute}}$. The submesh $S_{\text{protrusion}}(x, y, z)$ is formed from $\{v\}^i_{\text{impute}}$. We downsample $S_{\text{protrusion}}(x, y, z)$ by ¼ the number of vertex points and remesh using ACVD as described above both for computational efficiency and to get a higher quality mesh, $S^{ds}_{\text{protrusion}}(x, y, z)$ required for computing the intermediate equiareal disk parameterization for a final square parameterization. $S^{ds}_{\text{protrusion}}(x, y, z)$ is directly unwrapped to a 2D 128 x 128 pixel square image as described above. Positive curvature 'seed' regions are identified by thresholding the mean curvature mapped to 2D, $H\left(S^{ds}_{\text{protrusion}}(x, y, z)\right) > H_{thresh}$ with a global threshold and then applying morphological closing, disk kernel radius 1 pixel. To classify regions as having negative, flat and positive mean curvature, 3-class Otsu thresholding was applied to $H(S(x, y, z))$ to give two thresholds. All regions with mean curvature greater than the higher threshold $H_{thresh}$ were positive curvature. Undersegmented blebs correspond to a binary composed of conjoined pseudo-circular regions. We use the gradient watershed[114,115] on the Euclidean distance transform of the high curvature region binary to automatically separate conjoined blebs without seed markers. Mesh matching and interpolation was used to map $H$ and segmentation labels between $S_{\text{protrusion}}(x, y, z)$ and $S^{ds}_{\text{protrusion}}(x, y, z)$. The refined segmentation were mapped as seed labels from $S_{\text{protrusion}}(x, y, z)$ for every protrusion back to $S(x, y, z)$. The revised seed labels were then diffused across $S(x, y, z)$ using the combined geometrical and convexity affinity matrix from above for 10 iterations with $\alpha = 0.99$. The binary protrusion segmentation from above is applied, and any segmentation with Cartesian 3D surface area < 10 voxels$^2$ removed to give the final refined protrusion segmentation instances.

Topography guided decomposition of cell surface

The instance protrusion segmentation $F_{protrusions}(S(d, u, v))$ above assigns a unique protrusion label ID to each vertex of the surface mesh, $S(d, u, v)$. We use this surface-based $F_{protrusions}(S(d, u, v))$ as seed labels to partition the total internal cell volume into the volume space unique to each protrusion $i$, $V^i_{protrusion}(d, u, v)$ and the reference cortical cell volume, $V_{ref}(d, u, v)$. This is done in three parts; the construction of $V_{ref}(d, u, v)$, volume propagation of $F_{protrusions}(S(d, u, v))$, and using the previous two parts to volumize individual protrusions to obtain $V^i_{protrusion}(d, u, v)$ (Extended Fig. 4e-g).

*Construction of reference surface by imputation.* The reference surface with segmented protrusions removed, $S_{\text{ref}}(d, u, v)$ is of the functional form $S_{\text{ref}}(d = f(u, v), u, v)$ with $f(\cdot)$ injective and thus can be described by $d_{\text{ref}} = d = f(u, v)$ only. This is a 2D image with $d_{\text{ref}}$ as the pixel value. We impute the subset of pixels with missing values corresponding to the removed surface protrusions from pixels with known $d_{\text{ref}}$ using the fast marching image inpainting[116] implemented in the Python OpenCV library with an inpaint radius = 1 (Extended

Fig. 4e). The inpainted surface, $S_{\text{ref}}(d_{\text{ref}} = f_{\text{inpaint}}(u,v), u, v)$ is used to construct the binary reference cortical volume, $I_{binary}\left(V_{ref}(d,u,v)\right)$ which is 1 for all voxels whose $d < d_{\text{ref}} = f_{\text{inpaint}}(u,v)$.

*Volume propagation of surface-based instance protrusion segmentation.* The surface-based protrusion segmentation, $F_{protrusions}(S(d,u,v))$ is converted to voxel-based by setting the value of the voxels corresponding to the integer discretized $S(d,u,v)$ coordinates to the matching protrusion label ID. We expand the labels by 3 voxels using the Python Scikit-Image *skimage.segmentation.expand_labels* function and mask with the topographic binary cell volume $I_{binary}(V(d,u,v))$ to get the initial topographic volume protrusion segmentation, $I_{protrusions}(V(d,u,v))$ with only the surface of protrusions labelled. We apply marker watershed segmentation slice-by-slice to propagate labels laterally into the protrusion volume within a slice and labels from previous slices, from the top, $d = +D_{out}$ to the bottom, $d = -D_{in}$ of the topographic volume. At a slice $d = d_0$, we use the Euclidean distance transform of $I_{binary}(V(d = d_0, u, v))$ for watershed with the seed markers given by the labels of the previous slice, $d = d_0 + 1$ combined with the current labels of $I_{protrusions}(V(d = d_0, u, v))$ at slice $d = d_0$. In combining labels, the labels of the previous slice $d = d_0 + 1$ takes precedence and overwrites the label of $I_{protrusions}(V(d = d_0, u, v))$. The result, $I_{protrusions}(V(d,u,v))$ assigns a protrusion ID to all voxels in the entire cell volume $I_{binary}(V(d,u,v))$ (Extended Fig. 4f).

*Volumization of individual protrusions.* The binary reference cortical volume, $I_{binary}\left(V_{ref}(d,u,v)\right)$ is applied to exclude all cortical volume voxels in the watershed depth propagated $I_{protrusions}(V(d,u,v))$. We then apply connected component analysis to each unique protrusion label in $I_{protrusions}(V(d,u,v))$ and keep for each label, the largest contiguous volume region. The resulting $I_{protrusions}(V(d,u,v))$ is the final volume segmentation of all individual protrusions. For each unique protrusion, we generate a closed surface mesh by marching cubes. If the marching cubes mesh has > 1000 vertices, we downsample the mesh by a factor of 4 and remesh with ACVD. This last step is to keep the combined number of vertices across all protrusions and the reference surface reasonable for rendering and processing.

Direct Cartesian 3D decomposition of cell surface
For each segmented protrusion $i$, we construct the Cartesian 3D submesh, $S^i_{protrusion}(x,y,z)$. We find the set of vertices on the open boundary, $\{v\}_{boundary}$ using the Python libigl library function, *igl.boundary_loop*, compute the mean of these points, $\bar{v}_{boundary}$ and form a submesh, $S_{cap}(x,y,z)$ with $\{v\}_{boundary}$ and $\bar{v}_{boundary}$. We upsample $S_{cap}$ by successive mesh subdivision 3 times, each time replacing a triangle face by the four new faces formed by adding vertices at the midpoint of every edge, giving a mesh with $\approx 4^3 = 64$ times more vertices. Finally we solve the Possion problem[72] to find the vertex coordinates of $S_{cap}$ corresponding to the least bending energy. $S^i_{protrusion}(x,y,z)$ and $S_{cap}(x,y,z)$ are merged to form a closed surface mesh of protrusion $i$. Similarly $S_{cap}(x,y,z)$ is merged with the residual reference surface with segmented protrusions removed, $S_{\text{ref}}(x,y,z)$ to impute and close the hole left by protrusion $i$.

Conformalized mean curvature flow (cMCF) for flattening topographic surfaces, $S(d,u,v)$
$S(d,u,v)$ are open surfaces. Application of cMCF[49], which is designed for closed surfaces, maps $S(d,u,v)$ onto the 2D plane as an elliptical disk and in the limit to a point. We want the flow to converge to the planar $(u,v)$ rectangle. To do so, we impose additional no-flux constraints in the $u$-, $v$- directions on the boundary, $\partial S$ but allow flow in the depth, $d$ direction by adding to the right hand side of the cMCF equation an external force term that applies only in the $u$-, $v$- directions. In interior vertices, the flow follows the standard cMCF.

$$cMCF_{topo} := \begin{cases} \left(\mathbf{M}_t^{\text{boundary}} - \delta \mathbf{L}_0^{\text{boundary}}\right) v(t + \delta t) = \mathbf{M}_t^{\text{boundary}} v(t) + \underbrace{\left[\mathbf{0}_d | \left(-\delta \mathbf{L}_0^{\text{boundary}} v(t)\right)_{uv}\right]}_{\text{to ensure no flux in u,v direction}}, & \text{on } \partial S \\ \left(\mathbf{M}_t^{\text{mesh}} - \delta \mathbf{L}_0^{\text{mesh}}\right) v(t + \delta t) = \mathbf{M}_t^{\text{mesh}} v(t), & \text{on } S \setminus \partial S \end{cases}$$

where on the boundary, $\partial S$ we use the mass, $\mathbf{M}_t^{\text{boundary}}$ and Laplacian, $\mathbf{L}_0^{\text{boundary}}$ matrix defined for a 2D line and $\mathbf{M}_t^{\text{mesh}}$, $\mathbf{L}_0^{\text{mesh}}$ is the mass and Laplacian matrices defined for a 3D triangle mesh. [**A**|**B**] is used to denote the augmented matrix formed by appending the columns of matrix **A** and **B**. We solve for the vertex position at the next timepoint $v(t + \delta t)$ as with cMCF by direct matrix inversion.

## Surface curvature measurement

The mean curvature, $H$ of a 3D surface was measured as the divergence of $\hat{n}$, the unit surface normal[103], $H = -\frac{1}{2} \nabla \cdot \hat{n}$. The surface mesh is voxelized to a binary volume, $B$ and $\hat{n}$ is computed as the gradient of the signed distance transform of $B$ with the Euclidean distance metric. $H$ computed in this manner as opposed to from the mesh directly using discrete differential geometry[103] or quadric plane fitting[117] which is less affected by the number of mesh vertices or the mesh quality.

## Mesh quality measurement

The radius ratio $= 2 \frac{r_{in}}{r_{circ}}$, defined as twice the ratio between inradius and circumradius was used to measure the face quality for a triangle mesh in Extended. Fig. 2,3. It is a mesh quality measure in the sense that the radius ratio obtains its maximum value of 1 for an equilateral triangle; the shape which jointly maximizes all internal angles and gives the best conditioning number for the mesh Laplacian matrix[118].

## Surface rendering

Triangle meshes were exported from Python using the Python Trimesh library into .obj mesh files and visualized in MeshLab[119]. Volumetric images were rendered in Fiji ImageJ through the volume viewer plugin, and intensities were contrast enhanced for inclusion in the figures using Microsoft PowerPoint. The local surface maximum intensity projection image of Fig. 4c was produced by extending z-axis (depth) rays at every xy pixel, and taking the maximum intensity of voxels within $\pm 9$ voxels ($\pm 1 \mu m$) of the cell surface.

## Datasets

### Cell morphology validation dataset

To validate u-Unwrap3D (Fig. 2,3 and Extended Fig. 2-4), we used 66 cell surfaces segmented and surface protrusions classified using u-Shape3D and acquired from high resolution light sheet microscopy[4,120] as previously described[45]. The surfaces include 19 MV3 melanoma cells expressing Lifeact-GFP showing blebs, 38 dendritic cells expressing Lifeact-GFP showing lamellipodia, and 9 human bronchial epithelial (HBEC) cells expressing Tractin-GFP. We applied u-Unwrap 3D to these datasets with the following parameters for each step: for Step 1, cMCF with maximum iterations = 50, $\delta_t = 5 \times 10^{-4}$, stopping threshold, $\Delta_{thresh} = 1 \times 10^{-5}$ for blebs, $= 1 \times 10^{-5}$ for lamellipodia, $= 1 \times 10^{-4}$ for filopodia, $S_{\text{ref}}(x,y,z)$ mesh voxelization with morphological dilation and erosion with ball kernel radius 5 voxels, Gaussian smoothing $\sigma = 1$ of binary volume and initial $S_{\text{ref}}(x,y,z)$ meshing with marching cubes at isovalue 0.5, ACVD remeshing with number of clusters = 90% the number of vertices in the marching cubes mesh; for Step 3, area distortion relaxation with maximum iterations = 100, $\delta_t = 0.1$, stepsize $\varepsilon = 1$ and if equiareal was not achieved, repeat relaxation with a slower $\delta = 5 \times 10^{-3}$; for Step 4, the mean curvature of $S_{\text{ref}}(x,y,z)$ was used as the weight for determining the unwrapping axis, and a 1024 x 512 pixel $(u,v)$ grid; for Step 5, an upsampling factor of 3 for binary voxelization, $\alpha = 0.5$ voxel steps, a specified $D_{in} = 40$ steps and 2D robust smoothing[121] with smoothing factor = 50 for each iteration; for Step 6, binarization of the topographic 3D mapped binary cell segmentation with a threshold of 0.5, then Gaussian smoothing $\sigma = 1$ and initial marching cubes meshing at isovalue 0.5, ACVD remeshing with number of clusters = 50% the number of vertices in the marching cubes mesh.

### 3D timelapse lightsheet imaging and analysis of blebs

*Cell culture and timelapse imaging.* All details of the cell line creation, culture and imaging of the MV3 GFP-expressing melanoma cell movie in Fig. 4 were previously published[38]. The movie is a total of 200 frames acquired at a frequency of 1.21 s per frame. Each frame is a 104 x 512 x 512 size 3D volume with a voxel resolution of 0.300 x 0.104 x 0.104 μm.

*Cell segmentation and surface meshing.* The 200 timepoints were spatiotemporally registered volumetrically to the first timepoint, $t = 0$ as previously described[38]. The cell surface at $t = 0$ was segmented using a multi-level method[122] involving local contrast enhancement, deconvolution and edge enhancement and surface meshed as described above to obtain $S^{t=0}(x, y, z)$. Images were deconvolved using the Wiener-Hunt deconvolution approach[123] with our previously published point-spread function[45]. The surface mesh at all subsequent timepoints, $S^t(x, y, z)$ were reconstructed using the non-rigid registration deformation field from volumetric registration[38]. The vertex Septin intensity was calculated by extending from the surface a trajectory to an absolute depth of 1 μm along the steepest gradient of the distance transform to the mesh surface, and assigning the 95th percentile of intensity sampled along that trajectory to the originating vertex to capture the systematically brightest accumulation of Septin signal in the cortical shell. The raw Septin intensity suffers decay from bleaching. We simultaneously normalized and corrected the vertex Septin intensity by computing a normalized Septin intensity as the raw intensity divided by the mean Septin intensity in the whole cell volume at each timepoint.

*u-Unwrap3D analysis.* We computed a mean surface mesh, $\bar{S}(x, y, z)$ from all $S^t(x, y, z)$ as the input surface to u-Unwrap3D. This was done by surface meshing the mean binary volume over all binary voxelizations of individual $S^t(x, y, z)$ at an isovalue of 0.5. u-Unwrap3D was applied to $\bar{S}\{(x, y, z)$ to create a common static $(d, u, v)$ coordinate space that all $S^t(x, y, z)$ is mapped to in Step 5 of u-Unwrap3D to generate $S^t(d, u, v)$. u-Unwrap3D was run with the same parameters for all steps as for blebs in the validation dataset, except for the following modifications: step 1, the same automatic stopping iteration number but +5 steps, and ACVD with 10% of the marching cubes mesh to get a smoother $S_{\text{ref}}(x, y, z)$; step 4, a smaller 512 x 256 pixel size $(u, v)$ grid and not using the unwrapping axis inferred by curvature-weighted PCA - this axis passed through a large bleb and affected tracking; step 5, $D_{in} = 96$ steps - a total of 5μm. Topographic cMCF with the robust mesh Laplacian[102], mollify factor = $1 \times 10^{-5}$, $\delta_t = 5 \times 10^4$ was applied to each $S^t(d, u, v)$ for 10 iterations to compute the corresponding $S^t(u, v)$.

*Bleb segmentation and tracking.* Blebs were segmented from $S(d, u, v)$ at every timepoint using the instance segmentation algorithm with refinement for undersegmented blebs as described above. In computing the binary protrusion segmentation we use a downsampling factor of 4 due to the smaller 512 x 256 pixel $(u, v)$ grid and diffuse the segmentation for 5 iterations as the blebs were smaller than the validation dataset. The segmented 512 x 256 $(u, v)$ bleb images, $F_{bleb}(S(u, v))$ were padded 50 pixels on all four sides respecting spherical topology. This is done by periodic padding along the $u$- axis. For the $v$- axis, we pad the top of the image by reflecting the pixels with respect to the first image row (i.e. all pixels in row 2 to row 51) and then flipping in the $u$- axis. Similarly, the bottom is padded by refecting the pixels with respect to the last image row (i.e. all pixels in row 2 to row 511) and then flipping in the $u$- axis. For each unique bleb in every timepoint, we computed the bounding box of the bleb given by top left, $(u_{min}, v_{min})$ and bottom right $(u_{max}, v_{max})$ coordinates. The bleb bounding boxes were tracked using an optical flow assisted bounding box tracker[77]. Boxes were linked over time into tracks using bipartite matching and the intersection over union (IoU>0.25 for valid match) of bounding boxes as the distance function between pairs. To handle large changes in box size, the matching between the current and next frame was carried out on the predicted bounding box coordinates by local optical flow[124]. Optical flow was computed using the mean curvature, $H(S(u, v))$ after rescaling $H(S(u, v))$ to be an 8-bit grayscale image using the global minimum and maximum curvature values over time. In case of temporary occlusion or missed segmentation, any non-matched blebs were propagated for up to 5 frames (6s) using the estimated optical flow before track termination. Tracks with > 5 frames (6s) and a mean positive curvature, $H > 0.1$ μm$^{-1}$ were retained as bleb tracks. The coordinates of retained tracks was corrected to account for the initial padding of 50 pixels. To remove erroneous and duplicated tracks, we uniquely match every segmented bleb in each timepoint to a track by IoU. For each track, we then computed the fraction of its lifetime that could be matched to a bleb and removed all tracks for which this proportion was < 50%. Lastly for each track we checked for sudden changes in the bounding box area, which was indicative of an erroneous bounding box in need of substitution by an inferred corrected bounding box. We applied this procedure to each track in order to construct the timeseries of the bounding box area over the track lifetime and compute a smooth reference timeseries using the central moving average with a window of 3 frames. The bounding box at a timepoint is erroneous if the instantaneous difference between the raw and smooth bounding box area > 500 pixel$^2$ (the mean $(u, v)$ bleb box area is 361 pixel$^2$). The coordinates of a corrected bounding box is inferred from non-erroneous bounding boxes by interpolation using a linear spline. The tracks that remained fully in-focus over its lifetime were retained for analysis.

*Bleb timeseries extraction.*

We detected $(u, v)$ pixels on blebs by labelling spatially contiguous areas of positive mean curvature based on 3-class Otsu thresholding defining positive, flat or negative curvature. The largest connected component within a bleb bounding box was defined as on-bleb and the remainder area within the bounding box as off-bleb. We extracted distortion-corrected average timeseries of bleb area, mean curvature and septin intensity, that is of a scalar quantity, $F$ by observing that the mean of $F$ over a Cartesian 3D surface area is equivalent to computing a weighted mean over the equivalent $(u, v)$ area, $\frac{\iint_S F(S(x,y,z))\, dS}{\iint_S dS} = \frac{\iint_S F(S(u,v))\, dA\, dudv}{\iint_S dA\, dudv}$. The weight, $dA$ is the magnitude of the differential area element, $dA = \left|\frac{\partial S}{\partial u} \times \frac{\partial S}{\partial v}\right|$ described above.

*Bleb event alignment.*
Individual blebbing events were detected within a track by applying peak finding after central moving averaging of bleb area timeseries with a window of 3 timepoints. A peak was defined as having a prominence > 0.5 and separated from a neighboring peak by at least 3 timepoints. Individual bleb event timeseries were constructed and temporally aligned using the detected timepoint of maximal bleb area as timepoint 0 and taking a window of 14 timepoints on either side (a total 29 timepoints, 35 s).

3D timelapse lightsheet imaging and analysis of ruffles

*Cell culture and timelapse lightsheet imaging.* SU.86.86 cells were purchased from American Type Culture Collection (CRL-1837). The cells were transfected with integrating lentiviral plasmids carrying genes for myristoylated CyOFP1 and Tractin-mEmerald. The cells were cultured in RPMI medium supplemented with 10% fetal bovine serum and 1% anti-anti (Gibco 15240062), at 37°C in a humidified incubator and 5% $CO_2$. SU.86.86 cells were imaged on fibronectin-coated coverslips on a custom axially swept light sheet microscope[125]. The microscope detection system comprises a 25X NA1.1 water immersion objective (Nikon, CFI75 Apo, MRD77220) and a 500mm tube lens. The illumination was done through a 28.6X NA0.66 water immersion objective (Special Optics, 54-10-7). The movie analysed in Fig.5 is a total of 30 frames acquired at a frequency of 2.27 s per frame. Each frame is a two-channel 151 x 1024 x 1024 size 3D volume with a voxel resolution of 0.300 x 0.104 x 0.104 μm.

*Cell segmentation and surface meshing.* All timepoints were rigid registered volumetrically to the first timepoint, $t = 0$ to compensate for drift. The CyOFP1 image was also rigid registered to the Tractin-mEmerald in each timepoint. For every timepoint, the volumetric image was segmented using a multi-level method[122] involving local contrast enhancement, deconvolution and edge enhancement and surface meshed as described above to obtain the surface mesh, $S(x, y, z)$. The vertex Tractin-mEmerald and CyOFP1 intensities were calculated by extending a trajectory to an absolute depth of 1 μm along the steepest gradient of the distance transform to the mesh surface, and taking the mean intensity along the trajectory.

*u-Unwrap3D analysis.* The first timepoint surface mesh was used as the input $S(x, y, z)$ to u-Unwrap3D to create a common static $(d, u, v)$ coordinate space that the surface meshes from all timepoints is mapped to in Step 5 of u-Unwrap3D to generate $S^t(d, u, v)$. We use Unwrap-3D with the following parameters for each step: Step 1, cMCF with maximum iterations = 50, $\delta_t = 1 \times 10^{-5}$, automatic stopping threshold, $\Delta_{thresh} = 5 \times 10^{-5}$, $S_{\text{ref}}(x, y, z)$ mesh voxelization with morphological dilation and erosion with ball kernel radius 5 voxels, Gaussian smoothing $\sigma = 1$ of the binary volume and initial marching cubes meshing at isovalue 0.5, ACVD remeshing with number of clusters = 10% the number of vertices in the marchin cubes mesh, and further volume constrained Laplacian mesh smoothing[126] with implicit time integration, time step 0.5 for 15 iterations; for Step 3 area distortion relaxation with maximum iterations = 100, $\delta_t = 0.1$, stepsize $\varepsilon = 1$; for Step 4 we use the binary positive curvature region of $S_{\text{ref}}(x, y, z)$ given by 3-class Otsu thresholding as the weight for determining the unwrapping axis, and use a 1025 x 512 pixel $(u, v)$ grid; for Step 5, for outwards propagation, an upsampling factor of 3 for binary voxelization, $\alpha = $ minimum of 0.5 and $\frac{1}{2}(\Delta u + \Delta v)$ voxel -steps where $\Delta u, \Delta v$ is the mean Cartesian 3D distance of traversing one pixel in $u, v$ directions and a separable 1D uniform box filter smoother with a window 5 pixels; for inwards propagation, we use active contour cMCF with $\delta_t = 5 \times 10^{-4}$, and robust mesh Laplacian[102], mollify factor $1 \times 10^{-5}$ for better numerical stability with the $(u, v)$ parameterized $S_{\text{ref}}(x, y, z)$ converted into a triangle mesh by triangulating the quadrilateral pixel connectivity and inserting additional triangles to 'stitch' the image boundaries into a spherical topology, (note the latter stitching requires an even number of columns after discounting that the last column is the same as the first column, hence a 1025 x 512 grid) and $D_{in}$ = (maximum internal distance transform value ) / $\alpha$ steps; for Step 6, marching cubes meshing of the topographic 3D mapped binary cell segmentation at isovalue 0.5 following Gaussian smoothing $\sigma = 1$, ACVD remesh with number of clusters = 50% the number of vertices in the marching cubes mesh and retaining the largest connected component mesh. Topographic cMCF with robust mesh Laplacian[102], mollify factor $= 1 \times 10^{-5}$, $\delta = 5 \times 10^4$ was applied to each $S(d, u, v)$ for 10 iterations to compute the corresponding $S(u, v)$.

*Optical flow ROI tracking.* We used motion sensing superpixels (MOSES)[83,84] in dense tracking mode which automatically monitors the spatial coverage of ROIs and introduces new ROIs dynamically to ensure uniform spatial tracking at every timepoint. We partitioned the image with an initial user-specified 1000 non-overlapping rectangular regions-of-interest (ROI). Each ROI was tracked over time by subsequently updating its centroid by the median optical flow[124] within the ROI. Optical flow was computed from the Traction/CyOFP1 (TC) signal after rescaling $\text{TC}(S(u,v))$ to be an 8-bit grayscale image using the video minimum and maximum TC values.

*ROI timeseries extraction and cross-correlation.* Distortion-corrected average timeseries of mean Tractin/CyOFP1 (TC) and $H$ were computed for a track using the same weighted mean as for blebs. A square bounding box of the mean MOSES ROI width centered at the track $(u, v)$ coordinate was used to sample the scalar values at each timepoint. The distortion-corrected timeseries can be treated as standard 1D timeseries. The 1D normalized cross-correlation was thus computed between the distortion-corrected TC and $H$ timeseries for individual tracks without modification. ROI cross-correlation curves were averaged at all time-lags to derive the mean and 95% confidence interval ROI cross-correlation curve. A deviation of the curve greater than the 95% confidence interval at a time lag of 0 indicated significant instantaneous correlation.

*Retrograde actin flow and mean ruffle travel speed.* Computing the speed histogram with 25 bins and speed range 0-10 μm/min showed a slow and fast population (Fig. 5d). We inferred the mean speed of the two populations as the two thresholds generated by 3-class Otsu thresholding. The lower and faster of the thresholds are the mean speed of retrograde actin flow and ruffles respectively.

*Cross-correlation and curvature relationship.* We computed the continuous relationship of mean curvature, $H$ and the lag 0 cross-correlation of TC and $H$ (Fig. 5e) over ROI tracks using kernel density. Gaussian kernel density with a bandwidth set by Scott's rule was used to derive the joint density distribution of $H$ and cross-correlation i.e. $p(X, Y)$, with $X$: $H$, $Y$: cross-correlation over the closed intervals $X \in [-0.2, 0.6]$ and $Y \in [-1, 1]$. The continuous relationship is then given by the marginal expectation with capital denoting the random variable and $\mathbb{E}[\cdot]$ the expectation operator, $\mathbb{E}[Y|X = x] = \int Y p(Y|X = x) dY = \int Y \frac{p(X,Y)}{p(X)} dY = \frac{\int Y p(X,Y) dY}{\int p(X,Y) dY}$ with standard deviation equivalently defined as the square root of the variance, $\mathbb{E}[(Y - \bar{Y})^2 | X = x] = \mathbb{E}[Y^2 | X = x] - \mathbb{E}[Y | X = x]^2$. The evaluation of the integrals uses 100 bins for both $H$ and cross-correlation.

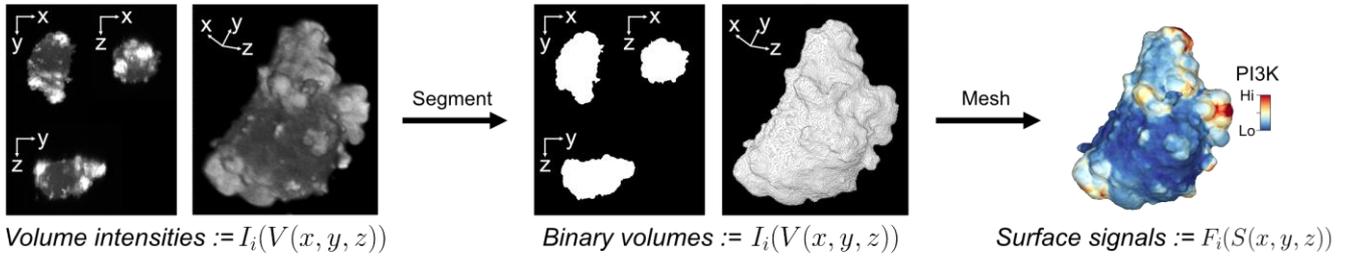
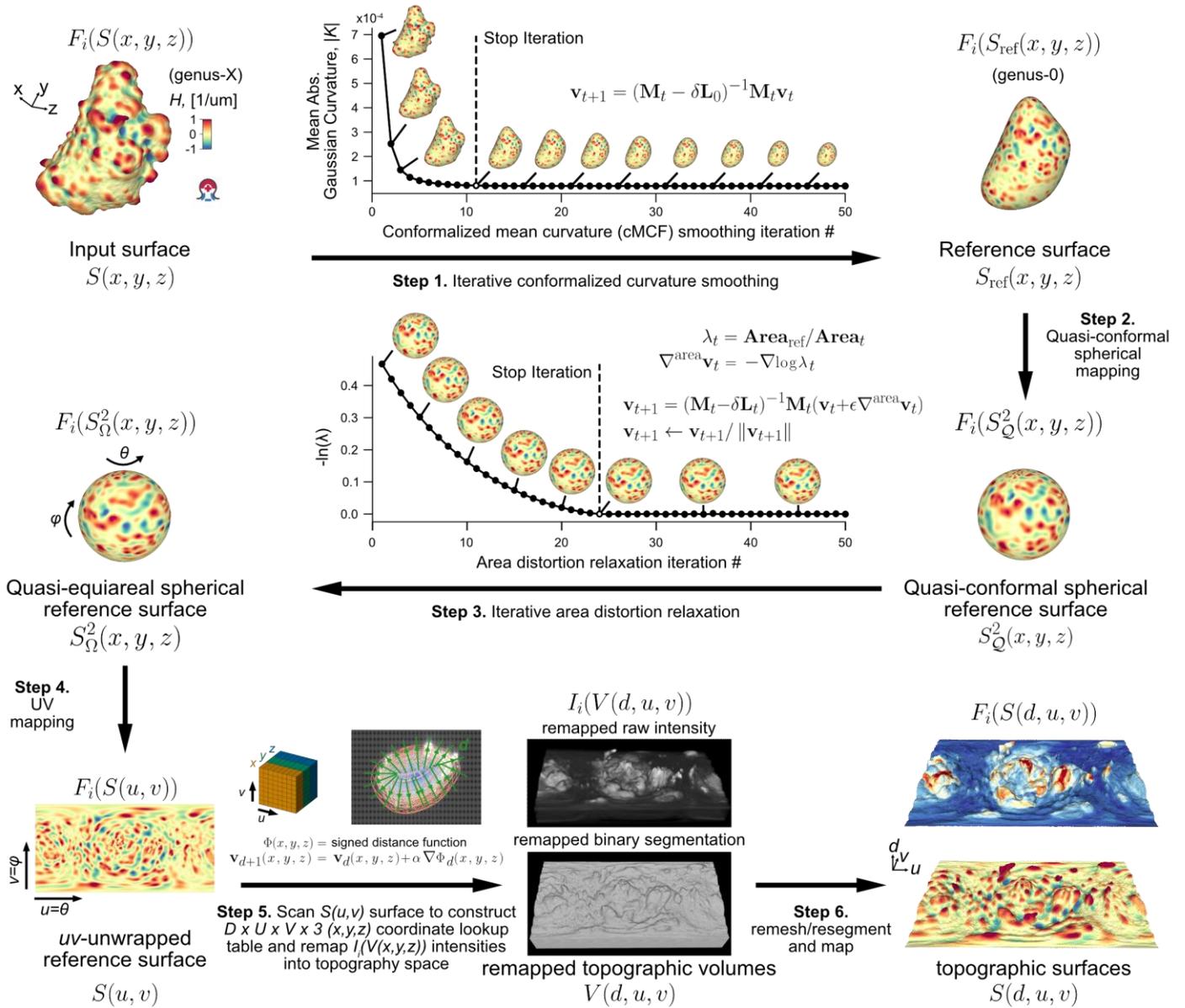
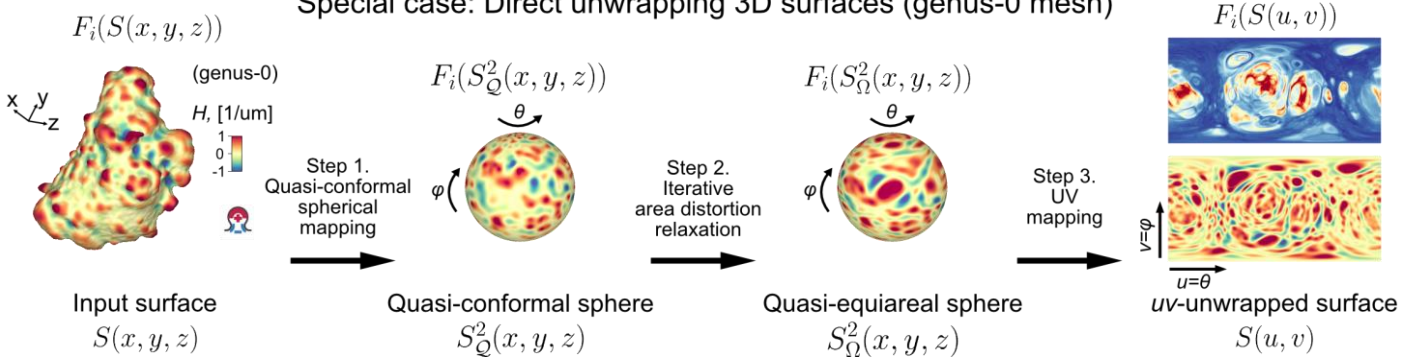

**Figure 1. Overview of the surface-guided computing framework u-Unwrap3D.** a) Overview of the 6 key steps to map an input genus-X Cartesian 3D surface, $S(x, y, z)$ such as that obtained from surface meshing the binary segmentation of an input 3D volume image, $I_i$, and associated scalar measurements, $F_i(S(x, y, z))$, via a smooth genus-0 reference surface, $S_{\text{ref}}(x, y, z)$, into any of three additional representations; topographic 3D surface, $S(d, u, v)$, 3D sphere, $S^2(x, y, z)$ and 2D plane, $S(u, v)$. **v** denotes mesh vertex coordinates, **M** is the mesh mass matrix, **L** is the mesh Laplacian matrix, $\epsilon$ the step size of area-distortion relaxation, $\alpha$ the step size (in pixels) of the propagation distance, $d$ the topographic depth (in pixels) and $t$ the iteration number. c) Steps to directly remap input genus-0 surfaces without need for a reference surface. In the figure, $S(\cdot), V(\cdot)$ denote surface and volume geometries, respectively, in either Cartesian 3D, topographic 3D, or radius-standardized 3D spherical coordinates; $F_i(S(\cdot))$ and $I_i(V(\cdot))$ denote surface or volumetric signals as a function of a particular surface or volume geometry. $H$ and $K$ denote mean and Gaussian curvatures respectively.

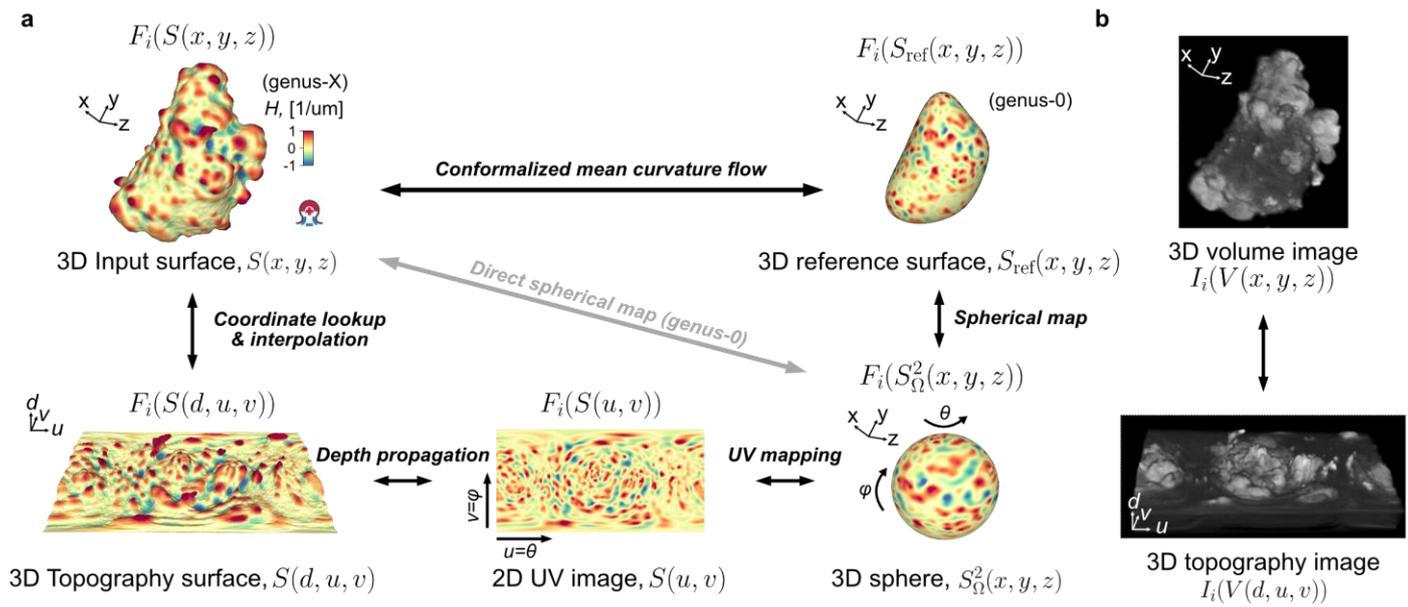

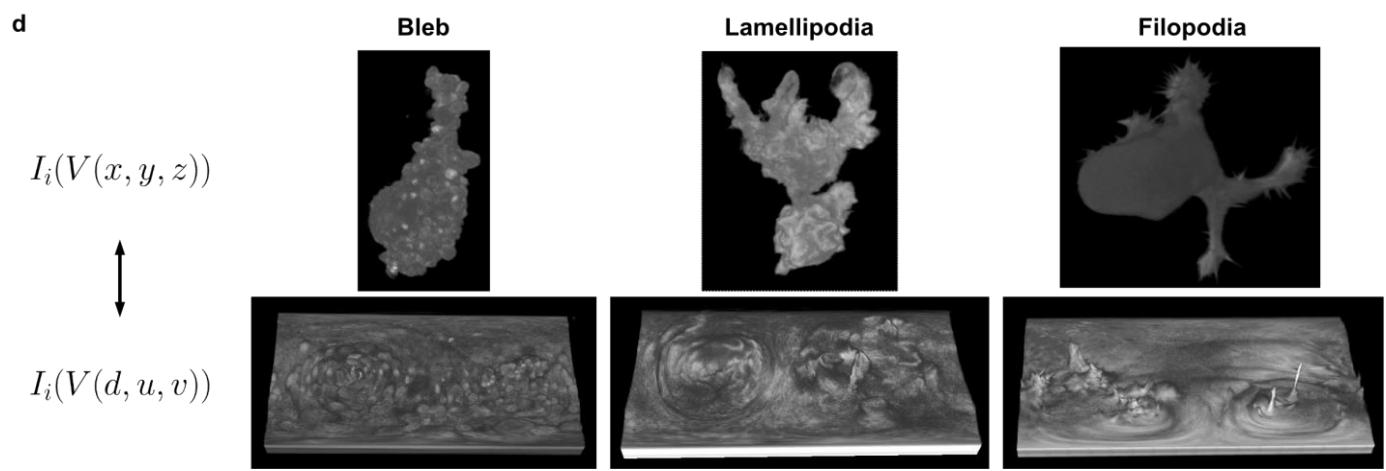

**Figure 2. u-Unwrap3D generates a spectrum of equivalent data representations for surface-guided computing. a)** Summary of the bijective mappings between the 5 equivalent surface mesh representations generated by u-Unwrap3D. Black bidirectional arrows indicate the mapping algorithms between representations discussed in the text. Grey arrow indicates the direct spherical mapping applicable when the input mesh is genus-0. **b)** u-Unwrap3D also enables bidirectional mapping of volumetric information between a Cartesian and topographic space relative to a genus-0 reference surface. **c)** Gallery of equivalent surface representations generated on examples of cell surfaces with blebs, lamellipodia and filopodia. For visualization of the mappings, individual instances of morphological motifs detected by the software uShape3D are color-coded on surface representations. **d)** Gallery of equivalent volume representations for the same cells shown in c). Volume image intensities were visualized using ImageJ volume viewer and contrast-enhanced to better visualize fine protrusions (see Methods).

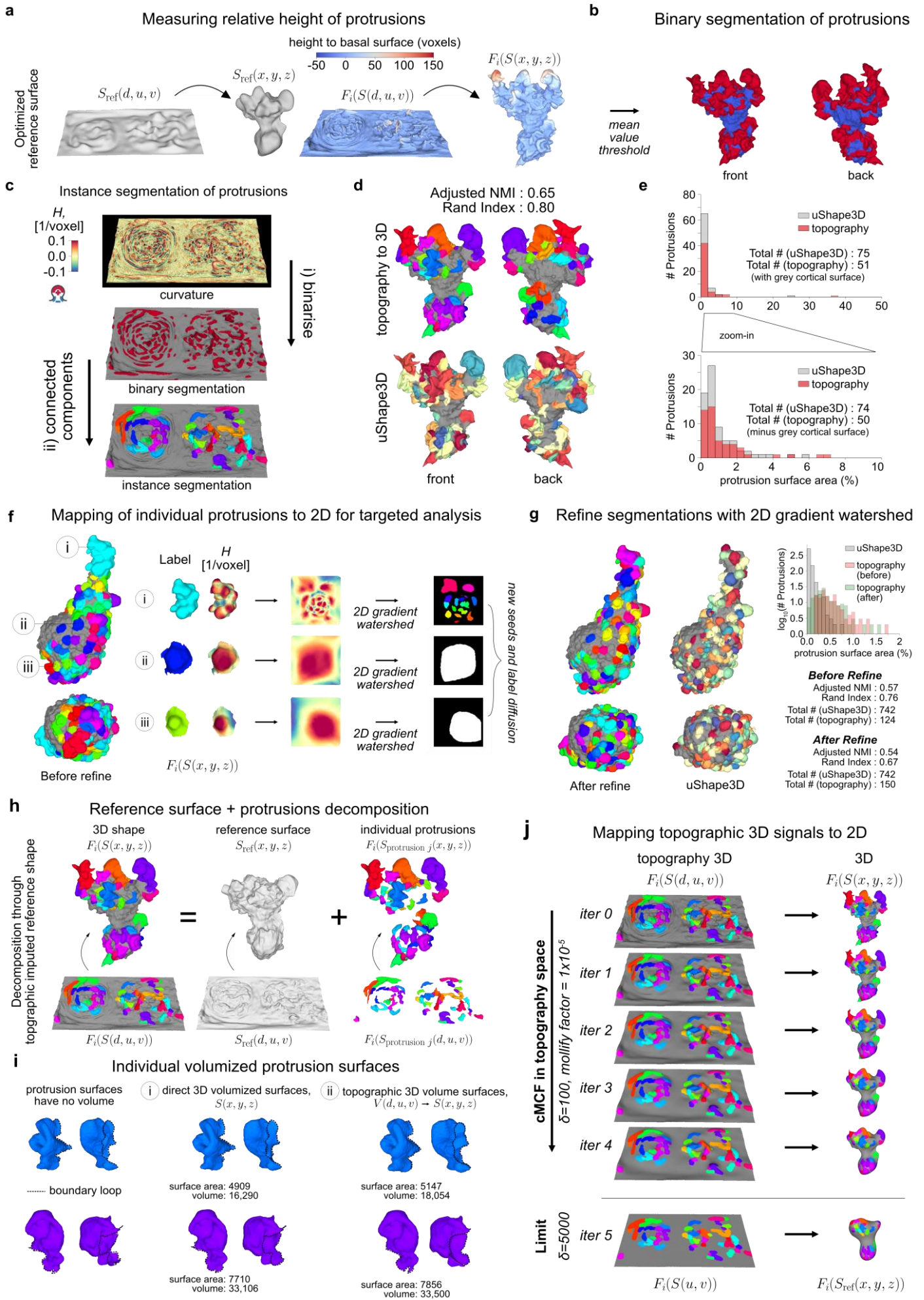

**Figure 3. u-Unwrap3D enables segmentation and characterization of complex 3D surface morphologies. a)** The $d$-coordinate of the topographic 3D surface directly measures the protrusion height $S(x, y, z)$ of the input surface relative to the reference surface $S_{\text{ref}}(x, y, z)$. Here, $S_{\text{ref}}(x, y, z)$ was optimized for delineating surface protrusions (see Extended Fig. 4a,b). **b)** Surface segmentation obtained by binary thresholding of the height measured relative to the optimized (blue) reference surface (see Extended Fig. 4c). Surface protrusions above the threshold are in red. **c)** Overview of an unsupervised pipeline to detect and segment protrusion instances by thresholding the topographic curvature and connected component labeling (see Extended Fig. 4d). Individual protrusions are uniquely colored. **d)** Comparison of the topography-guided protrusion segmentation with supervised uShape3D morphological motif detection. Individual protrusions are uniquely colored. Quantitative concordance was measured by adjusted normalized mutual information (NMI, 0-1) and Rand index (0-1). **e)** Comparison of the surface area of topography-guided (red bars) and uShape3D-based (grey bars) segmented protrusions plotted relative to the full reference surface (top) including the grey colored cortical surface and zoomed-in (indicated by the polygon) comparing only the surface area of segmented protrusions (bottom). **f)** Selective 2D unwrapping of 3 individual segmented protrusions labelled i-iii) into corresponding 2D disk and square representations for fine-grained segmentation of under-segmented protrusions **g)** Application of a watershed algorithm to the 2D representations refines under-segmented protrusions. The bijectivity of all intermediary mappings permits the representation of coarse- and fine-grained segmentations back on the 3D surface. Comparison of the final protrusion segmentation to the segmentation before refinement (see f)) and to the motifs detected by uShape3D. Quantitative concordance was measured by adjusted normalized mutual information (NMI, 0-1) and Rand index (0-1). **h)** Decomposition of an input Cartesian 3D surface (left) into reference cortical surface (grey colored) (middle) and individual meshes (uniquely colored) per segmented protrusion (right). The decompositions were guided by the topographic representations (bottom) (see Extended Fig. 4e-g). The decomposed surface meshes are closed and define individual volumes (i.e. volumized). **i)** Comparison of the surface area (left, labelled i) and volumes (middle, labelled ii) of individual protrusions computed from u-Unwrap3D topography guided from h) (x-axis) or standard 3D mesh processing (see Methods) volumized surface meshes (y-axis). Reconstructed reference cortical surface meshes without protrusions and colored by mean curvature using from u-Unwrap3D topography guidance (bottom) or standard 3D mesh processing (top) (right, labelled iii). Black triangles highlight mechanically implausible surface features left by standard 3D mesh processing. **j)** Illustration of the modified conformalized mean curvature flow (cMCF) to directly map topographic 3D surfaces and associated signals (here, segmented protrusions marked by unique colors) to the 2D plane, an optimal representation for tracking individual protrusions.

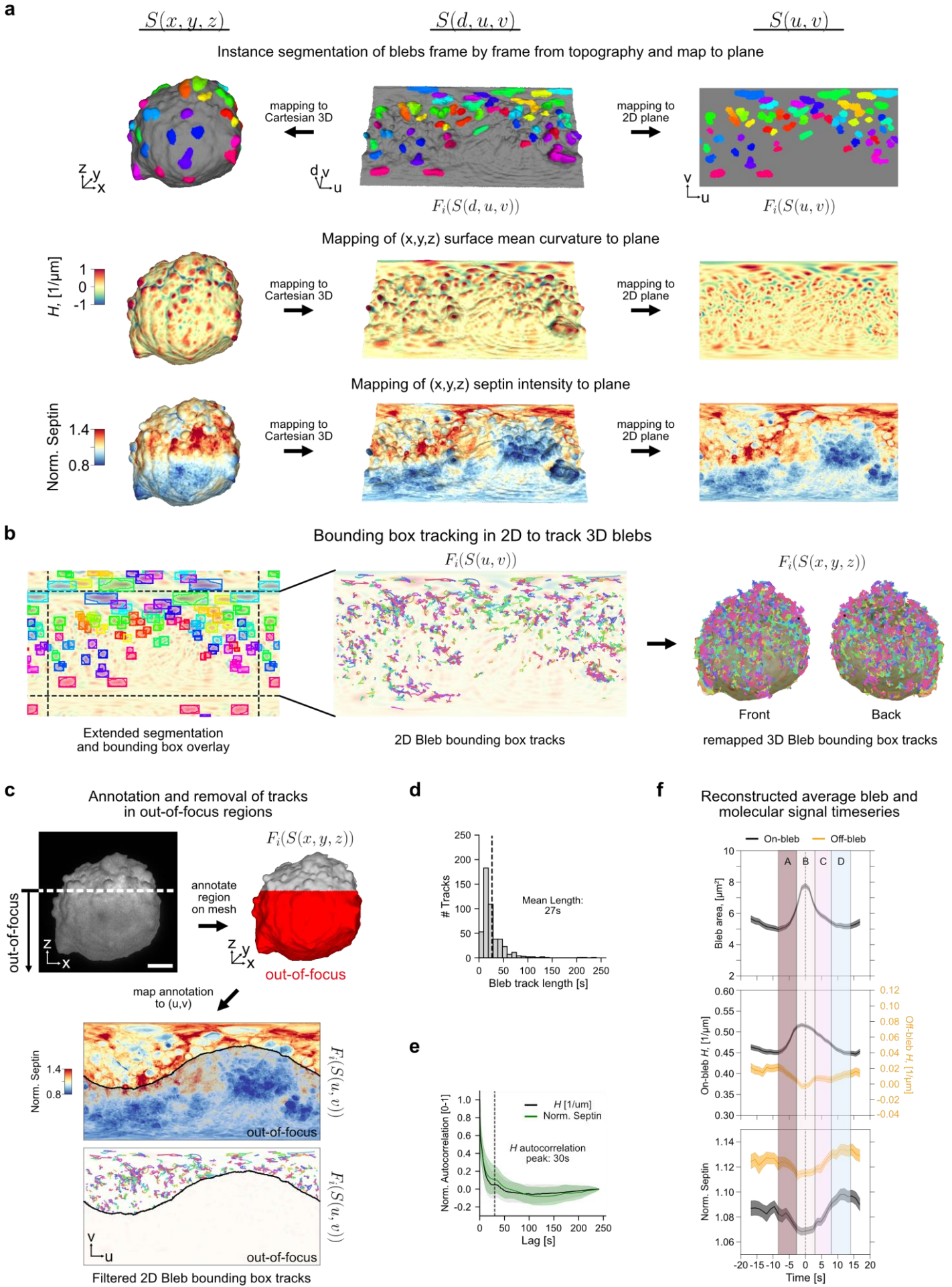

**Figure 4. u-Unwrap3D enables tracking and characterization of blebs and associated signals. a)** Individual blebs segmented in topographic 3D representation are mapped to Cartesian 3D for visualization and to the 2D plane for tracking (top). Individual blebs are uniquely colored. Mean curvature, $H$ (middle) and normalized Septin intensities (bottom) are jointly mapped from Cartesian 3D to topographic 3D to the 2D plane. The Septin intensity is normalized at each time to the mean Septin intensity in the whole cell volume to correct for expression variation and photobleaching. **b)** Tracking of individual blebs using an optical flow-guided 2D bounding box tracker. The unwrapped $(u, v)$-map is padded on all four sides to capture the continuation of the spherical surface (dashed black lines). Because of the bijectivity between representations individual bleb bounding box tracks in 2D (middle) can be mapped to 3D (right). **c)** Bijective mappings enable the transfer of manually annotated out-of-focus in Cartesian 3D to the unwrapped $(u, v)$ 2D plane to restrict intensity timeseries analyses to only the bleb tracks within the in-focus surface regions. The decay in image contrast with sample depth is shown in a maximum projection image of the first timepoint restricted to the segmented surface $\pm 1$ µm (Methods). Scalebar: 10 µm. **d)** Histogram of the in-focus bleb track lengths (dashed line, mean length). **e)** Autocorrelation curves (mean ± standard deviation) of mean curvature, $H$ and Septin computed from Cartesian 3D meshes. Dashed black line depicts the lag time of the first autocorrelation side lobe of mean curvature, $H$. **f)**. Average (mean ± s.e.m) time course of bleb surface area (top), mean curvature (on bleb, black; off bleb, orange, and Septin intensity (bottom) over a window 17.5s before to 17.5s after the timepoint of maximum bleb size used for alignment (n=545 bleb events from m=1 cell). A-D labels distinct phases of bleb-mediated curvature Septin recruitment.

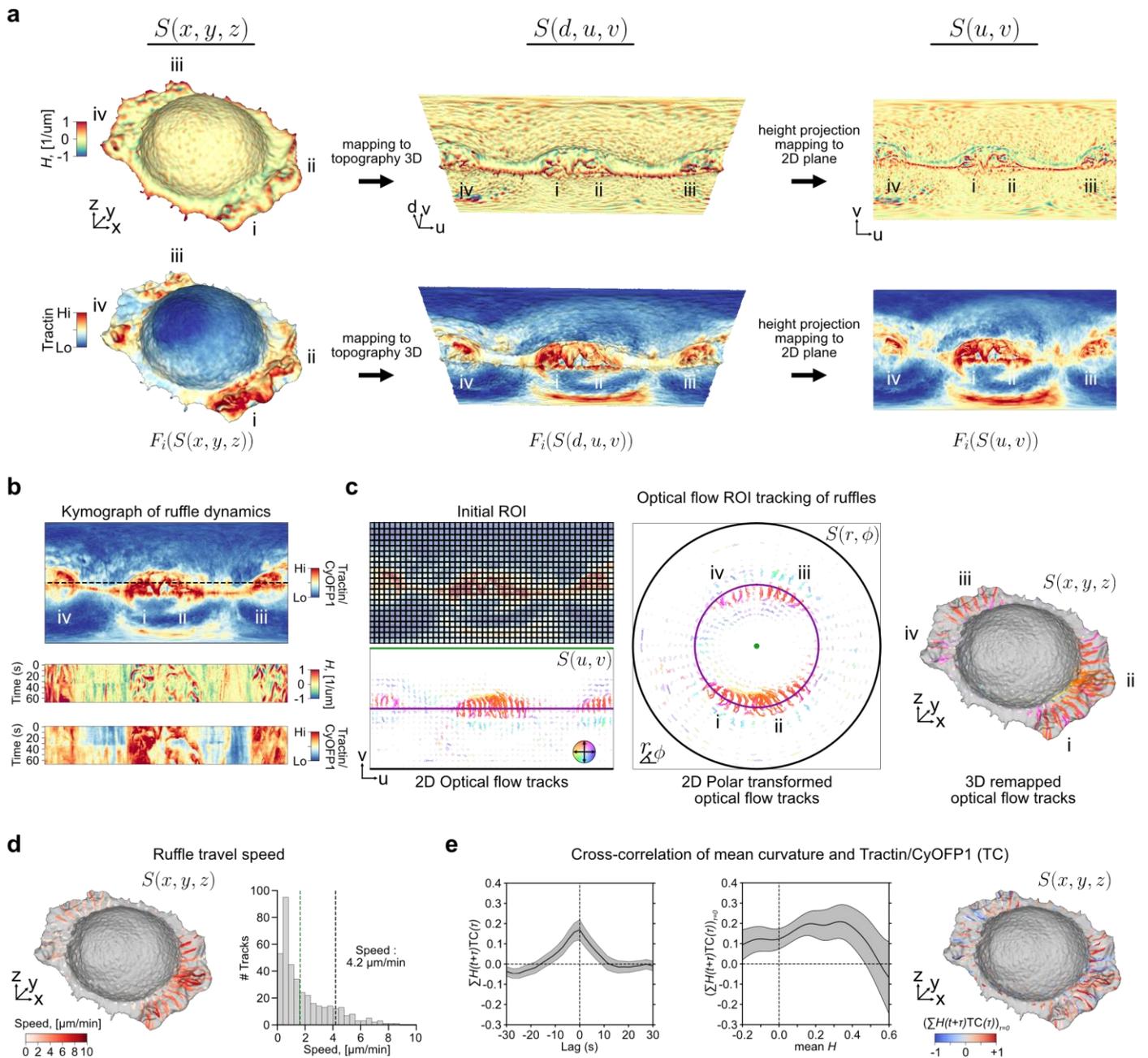

**Figure 5. u-Unwrap3D enables tracking and characterization of morphological and molecular dynamics. a)** Unwrapping of a ruffling SU.86.86 pancreatic adenocarcinoma cell visualized in Cartesian 3D (left), topographic 3D (middle) and unwrapped 2D plane (right) representations for the first frame of a timelapse 3D image sequence sampled every 2.27s for 30 frames. Top row: mean curvature; bottom row: Tractin-mEmerald intensity sampled 1 µm from the cell surface. Labels i-iv indicate corresponding landmarks in all three representations. **b)** Cross-section (dashed black line, top) to generate kymographs (bottom) of mean curvature and Tractin-mEmerald intensity normalized to myristolated CyOFP1 as a diffuse volumetric marker. **c)** Optical flow tracking on equipartitioned regions of interest (ROI) in the (u,v)-plane of ruffles based on the Tractin/CyOFP1 (TC) ratiometric intensity(left). The resultant optical flow ROI tracks are colored by the mean track direction. Color saturation indicates mean track speed. ROI tracks remapped to 2D polar $(r, \phi)$ view (middle) and to Cartesian 3D $(x, y, z)$ surface representation overlaid on the first time point (right). The polar transform maps the green (top), purple (middle) and black (bottom) horizontal line in the $(u,v)$-plane to the central green point, purple and black rings in the $(r, \phi)$-view, respectively. **d)** Mean temporal planar travel speed of the ruffle-associated ROI tracks from c) plotted onto the Cartesian 3D surface representation of the first time point (left) and histogram (right). We infer a mean ruffle travel speed of 4.2 µm/min corresponding to the faster of the two histogram populations (black vertical dashed lines) using 3-class Otsu thresholding (see Methods). **e)** Cross-correlation curve (mean ± 95% confidence interval) between mean curvature and TC timeseries per ROI track (left). Lag 0 cross-correlation of mean curvature and TC as a function of mean curvature, $H$ (middle); ROI tracks color-coded by cross-correlation magnitude plotted onto the Cartesian 3D surface representation of the first time point (right).

## a. Surface distorts under deformation

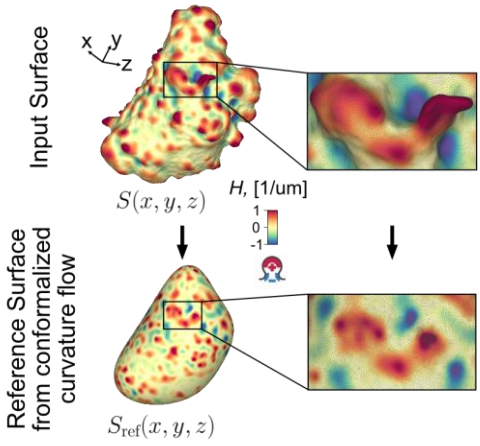

## b. Two types of deformation distortions

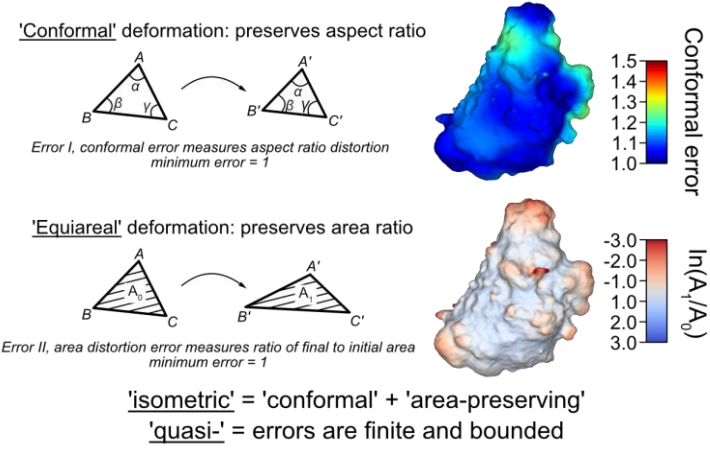

'Conformal' deformation: preserves aspect ratio

*Error I, conformal error measures aspect ratio distortion minimum error = 1*

'Equiareal' deformation: preserves area ratio

*Error II, area distortion error measures ratio of final to initial area minimum error = 1*

'isometric' = 'conformal' + 'area-preserving'
'quasi-' = errors are finite and bounded

## c. Conformalized mean curvature flow distortions

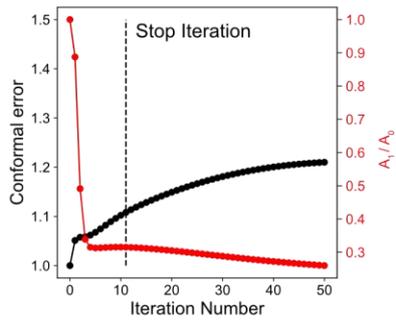

## d. Conformal spherical map distortions

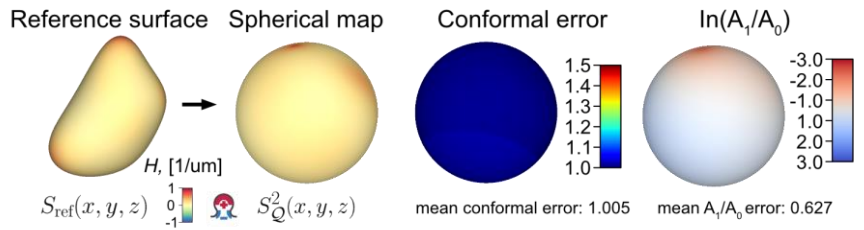

mean conformal error: 1.005    mean $A_1/A_0$ error: 0.627

## e. Area distortion relaxation distortions

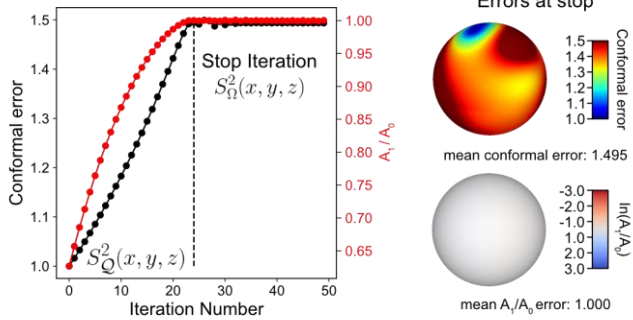

mean conformal error: 1.495
mean $A_1/A_0$ error: 1.000

## g. Surface to *uv* distortion

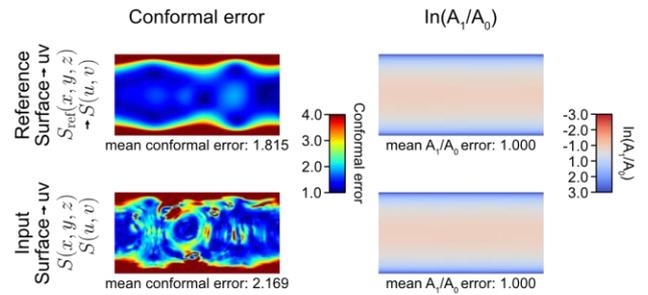

Reference Surface → uv $S_{ref}(x,y,z) \rightarrow S(u,v)$
mean conformal error: 1.815    mean $A_1/A_0$ error: 1.000

Input Surface → uv $S(x,y,z) \rightarrow S(u,v)$
mean conformal error: 2.169    mean $A_1/A_0$ error: 1.000

## f. Optimising (u,v) unwrapping axis

**Let:** point, $P_i = (x_i, y_i, z_i)$ on sphere
$w_i$ = associated weight of point, $P_i$ e.g. curvature

**Find:** optimal N-S axis, to visualise $w_i$ on (u,v) unwrapping

**Solution:** smallest eigenvector of matrix $A = (WP)(WP)^T$

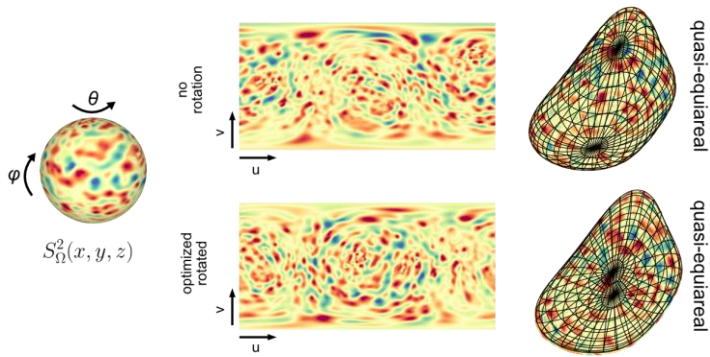

## h.

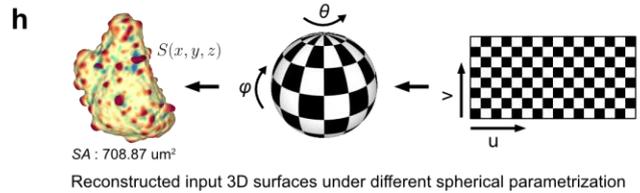

$SA$ : 708.87 um²

Reconstructed input 3D surfaces under different spherical parametrization

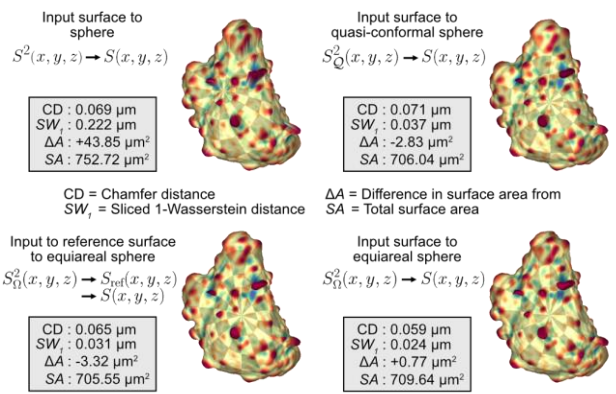

Input surface to sphere
$S^2(x,y,z) \rightarrow S(x,y,z)$
CD : 0.069 μm
$SW_1$ : 0.222 μm
$\Delta A$ : +43.85 μm²
$SA$ : 752.72 μm²

Input surface to quasi-conformal sphere
$S_Q^2(x,y,z) \rightarrow S(x,y,z)$
CD : 0.071 μm
$SW_1$ : 0.037 μm
$\Delta A$ : -2.83 μm²
$SA$ : 706.04 μm²

CD = Chamfer distance
$SW_1$ = Sliced 1-Wasserstein distance
$\Delta A$ = Difference in surface area from
$SA$ = Total surface area

Input to reference surface to equiareal sphere
$S_\Omega^2(x,y,z) \rightarrow S_{ref}(x,y,z) \rightarrow S(x,y,z)$
CD : 0.065 μm
$SW_1$ : 0.031 μm
$\Delta A$ : -3.32 μm²
$SA$ : 705.55 μm²

Input surface to equiareal sphere
$S_\Omega^2(x,y,z) \rightarrow S(x,y,z)$
CD : 0.059 μm
$SW_1$ : 0.024 μm
$\Delta A$ : +0.77 μm²
$SA$ : 709.64 μm²

**Extended Figure 1. Measuring and optimizing mesh distortion under surface deformation. a)** Any deformation of a closed 3D surface mesh (top) such as by conformalized mean curvature flow (cMCF) (bottom) distorts local geometrical distances and areas as illustrated by zoom-ins. **b)** Illustration of the two types of metric distortion incurred by mesh deformation; conformal and equiareal. In general, lower conformal error is at the expense of equiareal and vice versa. **c)** Plot of the conformal error (black dotted line, black left y-axis) and area ratio (red dotted line, red right y-axis) for each iteration of cMCF (Step 1, Fig. 1b) for the example mesh in a) and Fig. 1 with stop iteration indicated by a black vertical dashed line. **d)** Rendering of the conformal error and area ratio error at each triangle face for quasi-conformal spherical parametrization of the smooth shape to the sphere (Step 2, Fig. 1b). **e)** Plot of the conformal error (black dotted line, black left y-axis) and area ratio (red dotted line, red right y-axis) for each iteration of the spherical area distortion relaxation with stop iteration indicated by a black vertical dashed line (Step 3, Fig. 1b), (left). Rendering of the conformal and area ratio error of individual triangle faces at the stop iteration, (right). **f)** Illustration of not optimising (upper row) and optimising the unwrapping north-south axis using weighted principal component analysis to maximally display protrusive surface features with minimal distortion using the absolute value of mean curvature of the smooth 3D shape as weights, $w$ (lower row). **g)** Comparison of the per pixel conformal error (left column) and area distortion (right column) of the 2D $(u, v)$ unwrapping of the cMCF smooth shape, $S_{\text{ref}}(x, y, z)$ (upper row) or direct unwrapping of the input shape, $S(x, y, z)$ (lower row). **h)** Quantitative assessment of four different options of 2D $(u, v)$ unwrapping an input surface, $S(x, y, z)$ via different spherical parameterizations, $S^2(x, y, z)$ by measuring the difference between the Cartesian 3D reconstructed mesh from $S(u, v)$ and $S(x, y, z)$. $CD$ = Chamfer distance, $SW_1$ = sliced 1-Wasserstein distance between vertices of the input and reconstructed mesh. $SA$ = the total surface. $\Delta A$ = difference in total surface area between the input and reconstructed mesh. Qualitative assessment by uv-remapping the chessboard pattern and blending with the mean curvature, $H$ colors.

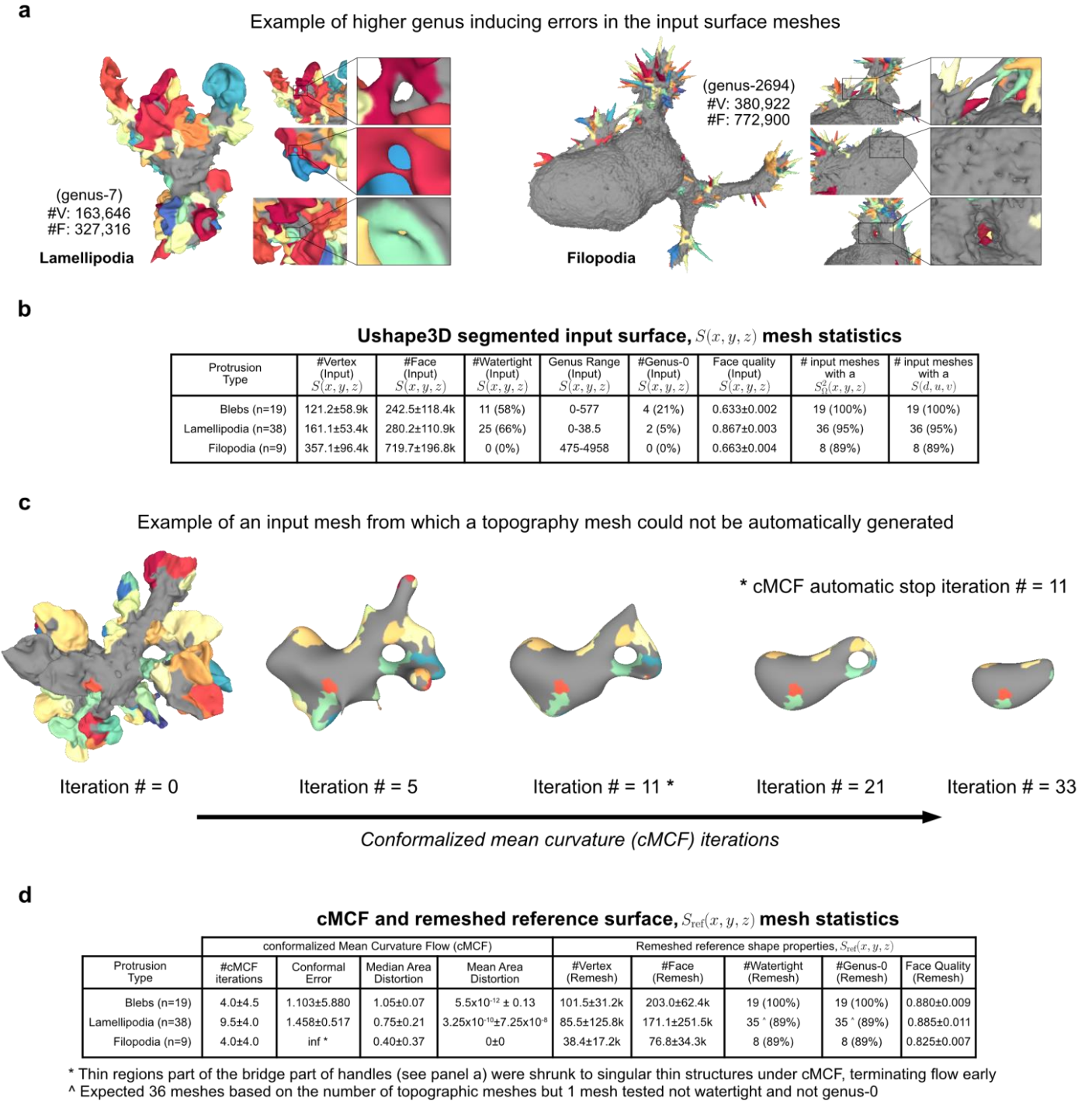

**Extended Figure 2. Assessment of the general applicability u-Unwrap3D. a)** Typical examples of higher genus inducing 'hole' and 'handle' errors in 3D meshes obtained by marching cubes meshing of binary segmentations from volumetric lightsheet microscopy images for two morphological motifs and two cell types; a dendritic cell with lamellipodia (left) and a HBEC cell with filopodia (right). **b)** Table summary of the statistics of the total 66 input surface meshes across morphological motif types to u-Unwrap3D (columns 1-6), of which for a total 63 input meshes (>95%) an equiareal spherical parametrization and topographic meshes were successfully computed (columns 7-8, last two columns). In comparison only 6 input meshes (9%) were genus-0. An example of a failed mesh is given in c). **c)** Example of u-Unwrap3D failure when the conformalized mean curvature flow (cMCF) reference shape at the automatic stop iteration has holes too large to be made genus-0 after voxelization and morphological hole closing in our current implementation (see Methods). **d)** Table summary of the conformal and area distortion error evaluated at the automatic stop iteration of the cMCF (left half, first 4 columns) and the summary mesh statistics of the remeshed cMCF smooth mesh (right half, last 5 columns). Note the area distortion is given as the surface area fraction ratio with the input surface $S_{\text{ref}}(x,y,z)$ as the denominator and equivalent to $1/\lambda$ used in the area distortion relaxation (see Methods). (see Methods). All tables in b)-g) report numerical values as median±interquartile range. An inf conformal error indicates local breakdown of flow for a mesh. When this occurs, cMCF cannot continue and the automatic stop iteration is the iteration # just prior to breakdown.

# a
**Quasi-conformal spherical parameterization, $S_Q^2(x,y,z)$ geometric distortion statistics**

| Protrusion Type | Conformal Error | Area Distortion |
|---|---|---|
| Blebs (n=19) | 1.002±0.003 | 0.978±0.037 |
| Lamellipodia (n=36) | 1.007±0.005 | 0.432±0.813 |
| Filopodia (n=8) | 1.016±0.013 | 0.140±0.580 |

# b
**Quasi-equiareal spherical parameterization, $S_\Omega^2(x,y,z)$ geometric distortion statistics for different stopping criteria**

| Protrusion Type | #iterations (equiareal) | Conformal Error (equiareal) | Area Distortion (equiareal) | Face Quality (equiareal) |
|---|---|---|---|---|
| Blebs (n=19) | 0.0±2.0 | 1.007±0.036 | 0.985±0.012 | 0.879±0.135 |
| Lamellipodia (n=36) | 23.0±13.0 | 1.602±0.946 | 1.000±0.000 | 0.771±0.149 |
| Filopodia (n=8) | 18.5±7.5 | 1.690±0.918 | 1.000±0.001 | 0.736±0.150 |

(i) Equiareal parameterization

| Protrusion Type | #iterations (MIPs) | Conformal Error (MIPs) | Area Distortion (MIPs) | Face Quality (MIPs) |
|---|---|---|---|---|
| Blebs (n=19) | 0±0 | 1.002±0.003 | 0.978±0.037 | 0.880±0.009 |
| Lamellipodia (n=36) | 0±0 | 1.007±0.005 | 0.432±0.813 | 0.885±0.010 |
| Filopodia (n=8) | 0±0 | 1.016±0.013 | 0.140±0.580 | 0.825±0.007 |

(ii) Most isometric parameterization (MIP)

| Protrusion Type | #iterations (Q+log$\lambda$) | Conformal Error (Q+log$\lambda$) | Area Distortion (Q+log$\lambda$) | Face Quality (Q+log$\lambda$) |
|---|---|---|---|---|
| Blebs (n=19) | 0.0±0.0 | 1.002±0.005 | 0.978±0.029 | 0.880±0.076 |
| Lamellipodia (n=36) | 7.5±14.8 | 1.254±0.594 | 0.866±0.112 | 0.846±0.095 |
| Filopodia (n=8) | 8.5±6.8 | 1.365±0.521 | 0.896±0.107 | 0.791±0.084 |

(iii) Conformal + equiareal parameterization

| Protrusion Type | #iterations (area-preserve MIPs) | Conformal Error (area-preserve MIPs) | Area Distortion (area-preserve MIPs) | Face Quality (area-preserve MIPs) |
|---|---|---|---|---|
| Blebs (n=19) | 4.0±3.5 | 1.073±0.056 | 0.997±0.011 | 0.877±0.130 |
| Lamellipodia (n=36) | 16.5±14.0 | 1.452±0.670 | 0.979±0.033 | 0.800±0.116 |
| Filopodia (n=8) | 12.5±4.5 | 1.550±0.715 | 0.980±0.028 | 0.759±0.122 |

(iv) Area-preserving most isometric parameterization (MIP)

# c
**Topographic 3D surface, $S(d,u,v)$ mesh statistics**

| Protrusion Type | #Vertex (Topographic) $S(d,u,v)$ | #Face (Topographic) $S(d,u,v)$ | Genus Range (Topographic) $S(d,u,v)$ | #Genus-0 (Topographic) $S(d,u,v)$ | Face Quality (Topographic) $S(d,u,v)$ |
|---|---|---|---|---|---|
| Blebs (n=19) | 389.3±23.9k | 776.2±47.8k | 0-10.5 | 17(89%) | 0.829±0.001 |
| Lamellipodia (n=36) | 535.0±138.2k | 1067.6±276.4k | 0-4.5 | 13(36%) | 0.831±0.002 |
| Filopodia (n=8) | 201.6±21.8k | 401.7±43.6k | 0-1.5 | 3(38%) | 0.873±0.003 |

# d
**Geometric error between Cartesian 3D remapped Topographic, $S_{\text{topo}}(x,y,z)$ and input, $S(x,y,z)$ surface mesh**

| Protrusion Type | Chamfer Distance, CD (voxel) | Sliced 1-Wasserstein, $SW_1$ (voxel) | % Difference in surface area, $\Delta A$ | % Difference in volume, $\Delta V$ |
|---|---|---|---|---|
| Blebs (n=19) | 2.79±6.72 | 4.28±7.24 | -11.6±43.9% * | +4.2±0.6% |
| Lamellipodia (n=36) | 1.77±0.16 | 0.93±0.33 | +1.2±5.4% | +7.9±1.4% |
| Filopodia (n=8) | 10.33±2.78 | 18.45±5.35 | -55.3±13.0% * | +3.4±1.2% |

* inflated error due to erroneous meshing of internal volume structures in input mesh

# e
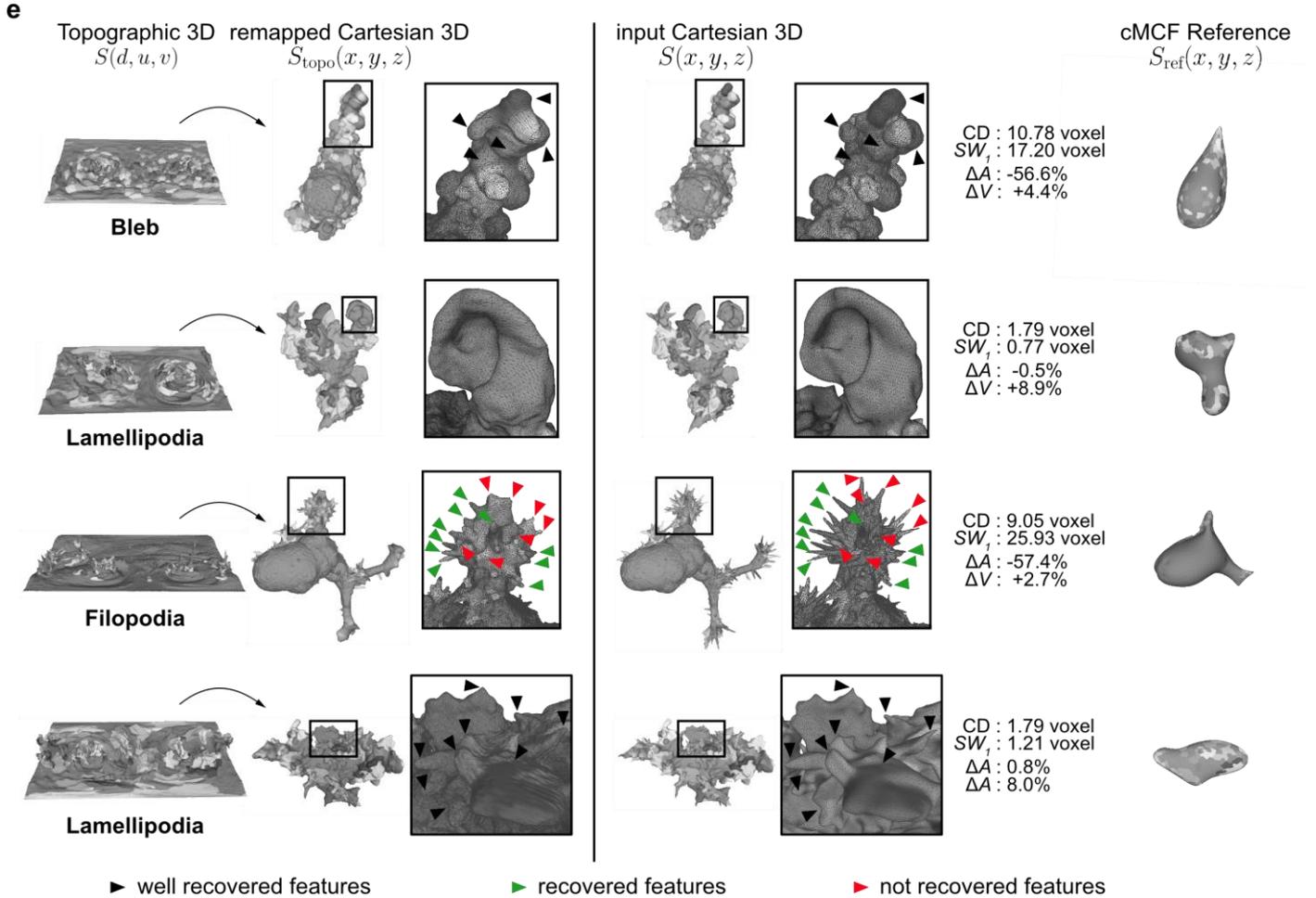

► well recovered features  ► recovered features  ► not recovered features

**Extended Figure 3. Quantitative performance assessment of the geometric deformation steps in u-Unwrap3D. a)** Table summary of the conformal and area distortion error of quasi-conformal spherical parametrization of the cMCF reference surface, $S_{\text{ref}}(x,y,z) \to S_Q^2(x,y,z)$, (Step 2 Fig. 1b). The target and optimal minimum conformal error is 1.0 and is achieved. **b)** Table summary of the number of iterations, conformal and area distortion error and mesh quality for four different stopping criteria (labelled i-iv, see Methods) for area-distortion relaxation of $S_Q^2(x,y,z) \rightarrow S_\Omega^2(x,y,z)$. The target and optimal minimum area distortion is 1.0. The implemented spherical area-distortion relaxation scheme in this paper achieves the optimal area distortion, an equiareal parametrization in $t_\Omega <$ a maximum allowed 50 iterations by minimising directly the area-distortion factor, $\lambda$ (equiareal) criterion (Methods) (top left, i). The scheme further allows relaxations between conformal and equiareal parametrizations as demonstrated by three additional stopping criteria: with no relaxation, (i.e. iteration 0) the parameterization is conformal and the most isometric parametrization (MIPS) (top right, ii); for iteration numbers $t \approx \frac{1}{2} t_\Omega$, the parameterization minimises jointly the combined conformal and area distortion as measured by $Q + \log \lambda$, the sum of the quasi-conformal error, $Q$ and the natural logarithm of the area distortion factor, $\lambda$ (bottom left, iii); and for iteration numbers $t \lesssim t_\Omega$, the parameterization is the area-preserving MIPS (bottom right, iv). **c)** Table summary of the statistics for computed topographic meshes, $S(d,u,v)$ using a 1024×512 pixel $(u,v)$ grid for the subset of $n=63$ meshes with successful equiareal spherical parametrizations, $S_\Omega^2(x,y,z)$ (Extended Fig. 2b). **d)** Table summary of the quantitative measurement of geometric error between the Cartesian 3D remapping, $S_{\text{topo}}(x,y,z)$ of the topographic mesh, $S(d,u,v)$ and the original input mesh $S(x,y,z)$ for four metrics; $(CD)$ chamfer distance (1st column), $(SW_1)$ sliced 1-Wasserstein (2nd column), $\Delta A$, the percentage difference in total surface area (3rd column) and $\Delta V$, the percentage difference in total volume (4th column). For a perfect reconstruction, all measures should be 0. Units are given as voxels due to heterogeneous pixel resolution amongst input meshes. A large $\Delta A$, but small $\Delta V$ for blebs and filopodia were due to a subset of non-watertight input meshes found to have erroneously meshed what should be the internal cell volume. These meshes were typically characterised by very high-genus (>50) (Extended Fig. 2c). Note in a)-d) the area distortion is given as the surface area fraction ratio with the input surface $S_{\text{ref}}(x,y,z)$ as the denominator and equivalent to $1/\lambda$ used in the area distortion relaxation (see Methods). All tables in a)-d) report numerical values as median±interquartile range. **e)** Quantitative and qualitative comparison of the Cartesian 3D remapping, $S_{\text{topo}}(x,y,z)$ of the topographic mesh, $S(d,u,v)$ (left) and the original input mesh $S(x,y,z)$ (middle) for 4 cell examples with different morphological motifs from d) in relation to the cMCF reference surface, $S_{\text{ref}}(x,y,z)$ (right). Box shows a zoom-in of the local surface region for each example. Black triangles highlight exemplar salient surface that are well captured but may be slightly smoothened and blurred in $S_{\text{topo}}(x,y,z)$ due to being underrepresented surface regions in $S(d,u,v)$, being distant from $S_{\text{ref}}(x,y,z)$. Green triangles highlight exemplar salient filopodia captured in $S_{\text{topo}}(x,y,z)$. Red triangles highlight exemplar salient filopodia not or poorly captured in $S_{\text{topo}}(x,y,z)$, due to being in a region of dense filopodia and is distant from $S_{\text{ref}}(x,y,z)$.

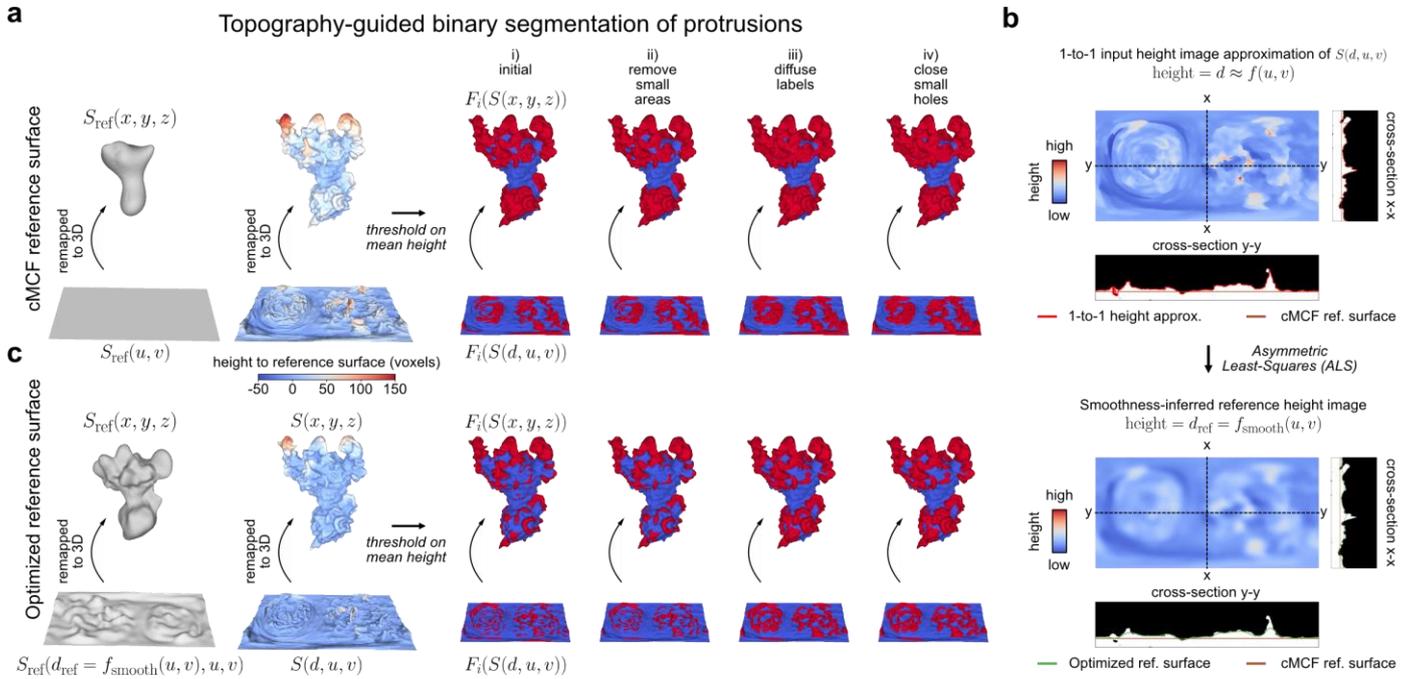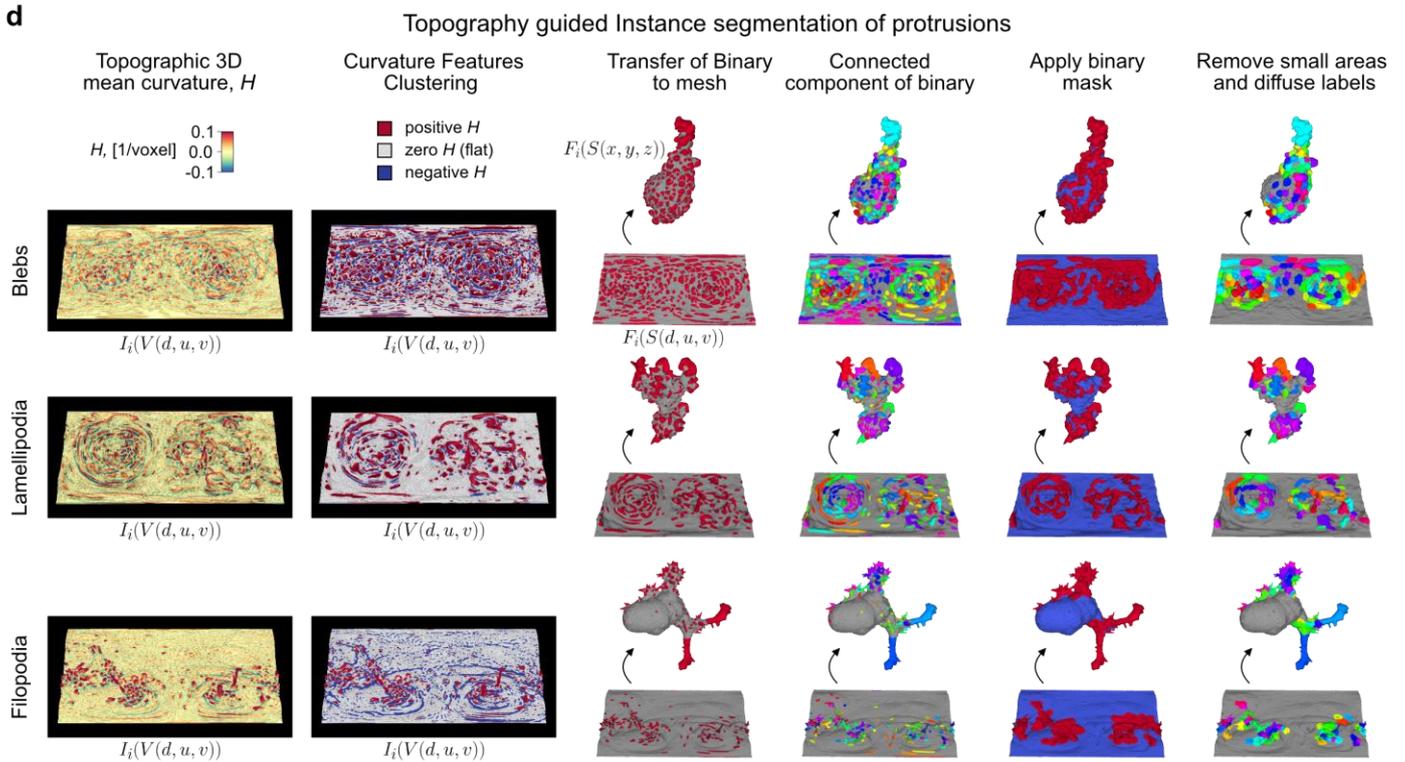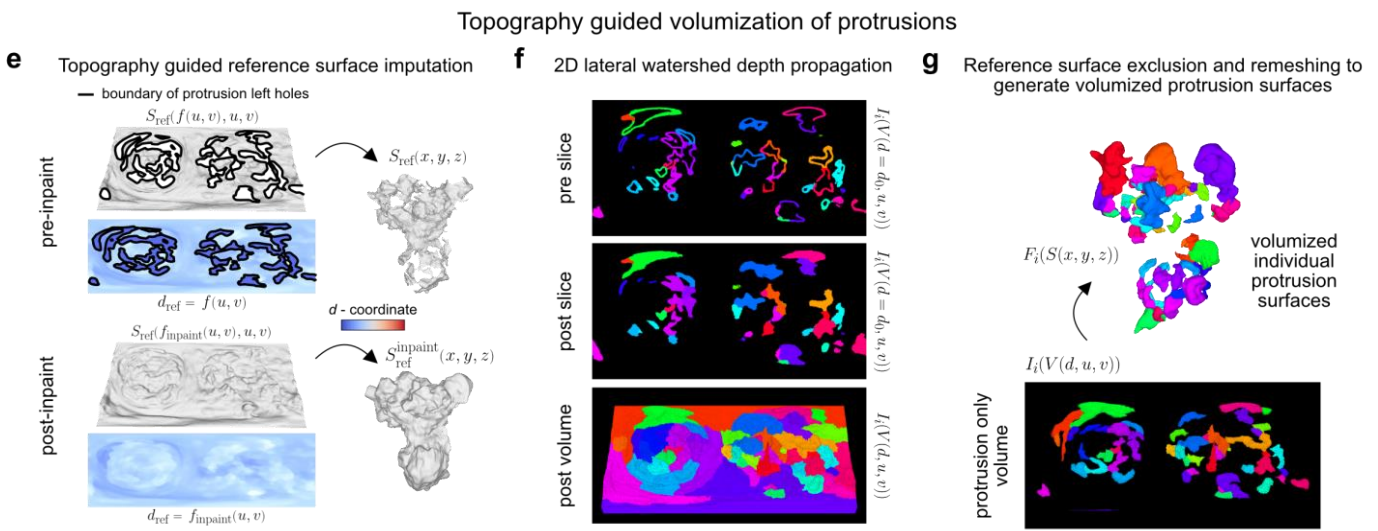

**Extended Figure 4. Overview of an unsupervised pipeline to segment complex surface morphologies guided by topographic 3D representations.**

**a)** Overview of the four key steps (i-iv) to binary segment protrusions by thresholding on the mean height measured relative to the cMCF reference surface. **b)** Learning an optimal reference surface for segmenting surface protrusions using asymmetric least mean squares (ALS) (Methods). The topographic surface, $S(d, u, v)$ is approximated by a surface, $S(d = f(u,v), u, v)$ that is in 1-to-1 correspondence to every $(u, v)$ pixel and can be represented as a height image using ray-propagation (top, see Methods). The height function, $f$ is depicted additionally in cross-section cuts x-x and y-y by a red line. A flat brown line in cross-section depicts the cMCF reference, $S_{\text{ref}}(d, u, v)$. ALS with smoothness regularization is applied to $d = f(u,v)$ to derive an optimal smooth reference surface with height, $d = f_{\text{smooth}}(u, v)$ (bottom). The height function, $f_{\text{smooth}}$ are depicted additionally in cross-section cuts x-x and y-y by a green line. **c)** Binary segmented protrusions by thresholding on the mean height measured relative to the ALS-derived reference surface from b). **d)** Overview of the sequential steps, left-to-right to segment individual protrusions by binarization and connected components analysis of topographic volume signals, $I_i(V(d, u, v))$. The steps are illustrated for 3 different cell types and 3 different surface motifs; MV3 melanoma cell with blebs (top row), dendritic cell with lamellipodia (middle row) and HBEC cell with filopodia (bottom row). Initial binarization uses 3-class k-means clustering (blebs and filopodia) and 3-class Gaussian mixture model clustering (lamellipodia) of volumetric mean curvature to identify all positive curvature regions. **e)** The hole-ridden ($d = 0$, black outline, and dark blue colored) reference surface, $S_{\text{ref}}(d_{\text{ref}} = f(u, v), u, v)$ (left) and corresponding remapped Cartesian 3D surface (right) after removal of all protrusion mesh faces (top). The inpainted topographic 3D reference surface, $S_{\text{ref}}(d_{\text{ref}} = f_{\text{inpaint}}(u, v), u, v)$ (left) and corresponding remapped Cartesian 3D surface (right) after image inpainting the missing $d$ coordinates (bottom). **f)** Marker-seeded 2D watershed to laterally propagate surface segmentation labels into the topographic volume shown pre- (top) and post- (middle) for a single given slice at depth, $d = d_0$. The resulting labelled topographic volume after propagating all surface protrusion labels fully from the top to the bottom (bottom). Unique colors denote unique regions with the same surface label. **g)** The protrusion-only topographic volume labels (bottom) and resultant composition of individual volumized protrusion meshes (top) after using the inpainted basal surface from e) to mask out all non-protrusion voxels in f) and meshing. Individual protrusions are uniquely colored.

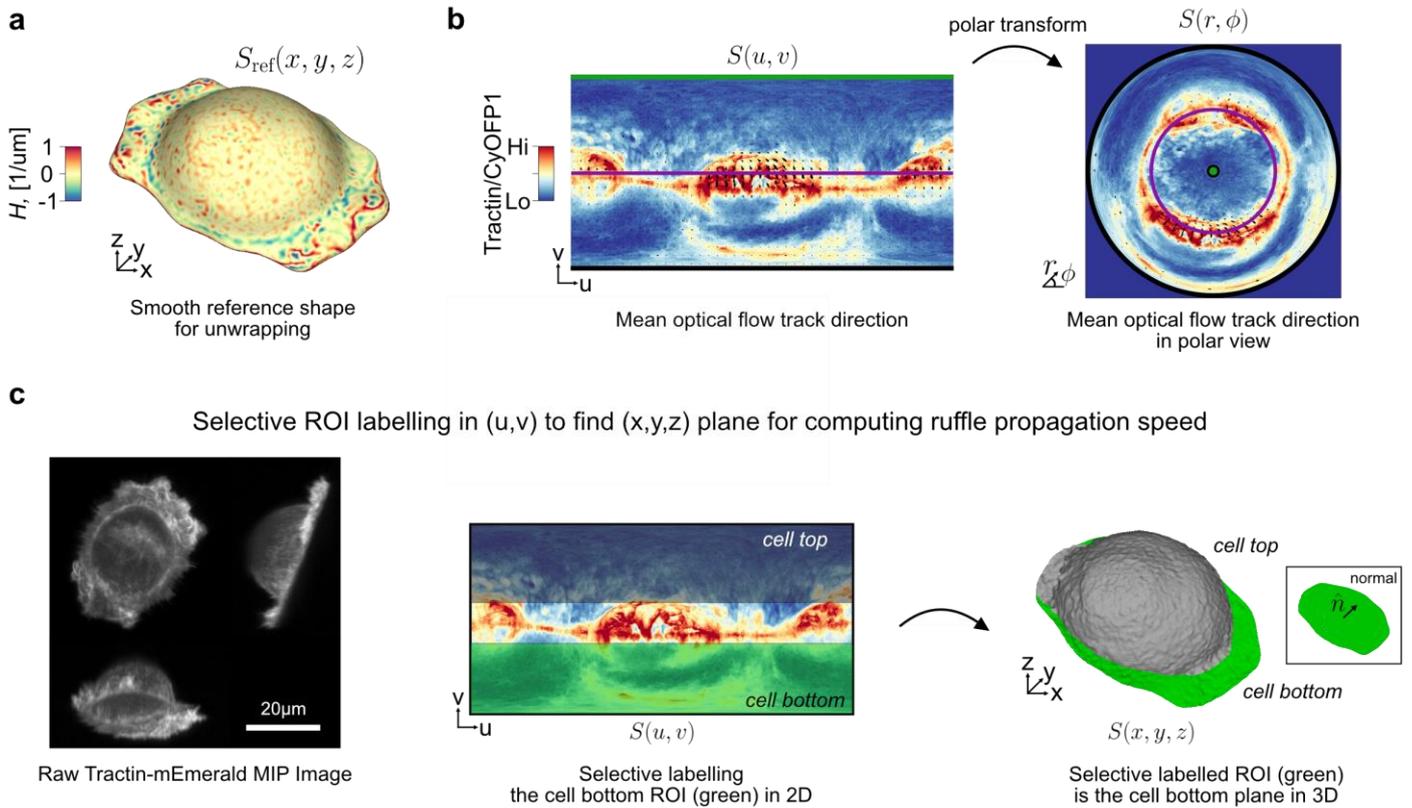

**Extended Figure 5. u-Unwrap3D enabled measurement of ruffle in-plane travel speed. a)** Cortical reference shape, $S_{\text{ref}}(x, y, z)$ found by applying conformalized mean curvature flow to the first timepoint of the ruffling SU.86.86 cell in Fig. 5a and used to unwrap all timepoints into a common static $(d, u, v)$ space. **b)** The mean track velocity of the optical flow region-of-interest (ROI) tracks, plotted as black arrows at the initial track coordinate overlaid on the unwrapped Tractin-mEmerald/CyOFP1 intensity image of Fig. 5b (left) and its corresponding polar transformed equivalent (right). The polar transform maps the green (top), purple (middle) and black (bottom) horizontal line on the left to the central green point, second purple and third black rings in the polar image. **c)** Orthogonal cross-section maximum intensity projection (MIP) image of the raw Tractin-mEmerald intensity channel showing the tilted lightsheet acquisition (left). Selective ROI isolation of the top and bottom surface of the cell by grey and green bounding box selection in unwrapped view (middle) and visualized in 3D with grey and green surfaces respectively (right). The normal vector, $\hat{n}$ describing the best fit plane through only the cell bottom vertices found by principal components analysis (inset rectangle).

# Supplementary Videos

**Supplementary Video 1.** Overview of the six key steps of u-Unwrap3D.

**Supplementary Video 2.** Application of u-Unwrap3D to directly unwrap a genus-0 cell surface mesh in conjunction with the spatial activation pattern of PI3K signaling products.

**Supplementary Video 3.** Application of u-Unwrap3D to an MV3 melanoma cell with bleb surface motifs segmented by the u-shape3D software. Each bleb is labelled with a random color to demonstrate the local surface mappings and their distortions.

**Supplementary Video 4.** Application of u-Unwrap3D to a dendritic cell with lamellipodia surface motifs segmented by the u-shape3D software. Each lamellipodium is labelled with a random color to demonstrate the local surface mappings and their distortions.

**Supplementary Video 5.** Application of u-Unwrap3D to an HBEC cell with filopodia surface motifs segmented by the u-shape3D software. Each filopodium is labelled with a random color to demonstrate the local surface mappings and their distortions.

**Supplementary Video 6.** Application of topographic conformalized mean curvature flow to directly map the topographic surface $S(d, u, v)$ of a dendritic cell with segmented lamellipodia surface motifs to the 2D plane, $S(u, v)$ for two different time steps, $\delta = 100$ and $\delta = 5000$. The smaller $\delta$ enables gradual relaxation and the ability to sample and use intermediate shapes during the flow. However smooth low curvature folds remain such that we do not fully converge to the plane even if continued to 100 iterations. For direct mapping to the plane we always use the large $\delta$ to ensure convergence within 50 iterations.

**Supplementary Video 7.** Application of u-Unwrap3D to enable segmentation and tracking of blebs on a MV3 melanoma cell in topographic representation. View 1: Projections of cell surface, mean curvature and normalized SEPT6-GFP into topographic surface and $(u, v)$ unwrapped reference surface representations. Individual blebs are segmented by thresholding in the topographic surface representation. Leveraging the bijectivity of u-Unwrap3D mappings bleb labels are projected back to the 3D surface. View 2: The 2D segmented blebs are tracked and trajectories projected back to the original 3D surface. The timelapse volumes were acquired every 1.21s for 200 frames. Scalebar: 10μm.

**Supplementary Video 8.** Application of u-Unwrap3D to track the surface ruffling and actin flows of a SU.86.86 pancreatic adenocarcinoma. View 1: Projections of cell surface, mean curvature and Tractin-mEmerald into topographic surface and $(u, v)$ unwrapped reference surface representations. View 2: Regional ruffling and actin flows are tracked in 2D with optical flow and trajectories projected into a 2D polar representation and back to the original 3D surface. View 3: Select measurement of instantaneous ruffle and actin flow speeds and cross-correlation of actin and curvature within the lamella and lamellipodia taking advantage of the unwrapped $(u, v)$ representation is projected back to the original 3D surface. The timelapse volumes were acquired every 2.27s for 30 frames. Scalebar: 20μm.